\documentclass[11pt,twocolumn]{article}

\usepackage{amsmath,amsfonts,amssymb,bbm,enumitem,url,color,graphicx,amsthm,rotating,authblk,soul,siunitx,tabulary,booktabs,hyperref}
\usepackage[font=footnotesize,labelfont=bf]{caption}
\usepackage[a4paper]{geometry}
\usepackage[round]{natbib}

\begin{document}
	
	\twocolumn[	\begin{@twocolumnfalse}
		\begin{center}
			
			{\bf \Large A systematic comparison of tropical waves over northern Africa. Part II: Dynamics and thermodynamics}
			
			\bigskip
			
			\medskip
			
			{\bf Andreas Schlueter\textsuperscript{*1}, Andreas H.~Fink\textsuperscript{1}, Peter Knippertz\textsuperscript{1}}
			
			\textsuperscript{1} Institute of Meteorology and Climate Research, Karlsruhe Institute of Technology, Germany 
			
			\textsuperscript{*}\url{andreas.schlueter@kit.edu} 
			
			\bigskip
			
			\textbf{This work has been submitted to Journal of Climate.\\Copyright in this work may be transferred without further notice.}
			
			\bigskip

		\end{center}

\abstract{This study presents the first systematic comparison of the (thermo-)\nolinebreak[0]dynamics associated with all major tropical wave types causing rainfall modulation over northern tropical Africa: Madden Julian Oscillation (MJO), Equatorial Rossby waves (ERs), mixed Rossby-gravity waves (MRGs), Kelvin waves, tropical disturbances (TDs, including African Easterly Waves), and eastward inertio-gravity waves (EIGs). Reanalysis and radiosonde data were analyzed for the period 1981--2013 based on space-time filtering of outgoing longwave radiation. The identified circulation patterns are largely consistent with theory. The slow modes, MJO and ER, mainly impact precipitable water, whereas the faster Kelvin waves, MRGs, and TDs primarily modulate moisture convergence. Monsoonal inflow intensifies during wet phases of the MJO, ERs, and MRGs, associated with a northward shift of the intertropical discontinuity for MJO and ERs. During passages of vertically tilted imbalanced wave modes, such as MJO, Kelvin waves, and TDs, and partly MRGs, increased vertical wind shear and improved conditions for up- and downdrafts facilitate the organization of convection. The balanced ERs are not tilted and rainfall is triggered by large-scale moistening and stratiform lifting. The MJO and ERs interact with intraseasonal variations of the Indian monsoon and extratropical Rossby wave trains. The latter causes a trough over the Atlas Mountains associated with a tropical plume and rainfall over the Sahara. Positive North Atlantic and Arctic Oscillation signals precede tropical plumes in case of the MJO. The results unveil which dynamical processes need to be modeled realistically to represent the coupling between tropical waves and rainfall in northern tropical Africa.\\	
	\bigskip}

\end{@twocolumnfalse}]

	\section{Introduction} 
	
	The West African monsoon (WAM) system strongly affects livelihoods in northern tropical Africa. The highly weather-dependent agricultural sector employs \SI{60}{\percent} of the work force in West Africa and contributes \SI{35}{\percent} of the GDP \citep{Jalloh.2013b}. As the WAM exhibits considerable variability on daily to subseasonal timescales \citep{Janicot.2011}, farmers and pastoralists can substantially benefit from weather forecasts and adapt their cropping \citep{Roudier.2016} or herding strategies \citep{Rasmussen.2014}. However, global numerical weather prediction (NWP) models almost entirely fail so far to deliver accurate and reliable weather forecasts for this region \citep{Vogel.2018}. 
	
	Although rainfall over northern tropical Africa is highly stochastic by nature, predictability exists on different temporal and spatial scales, as, e.g., tropical waves are known to modulate precipitation. Specifically, convectively coupled equatorial waves (CCEWs) stand out against the background noise of tropical rainfall \citep{Wheeler.1999}. The theory of a shallow water system yields several eigenmodes of motion that are trapped near the equator \citep{Matsuno.1966}. These waves can broadly be classified into balanced and imbalanced modes. The former are called equatorial Rossby (ER) waves, because they follow quasi-geostrophy, just as Rossby waves in the mid-latitudes. The latter are inertio-gravity waves, which are subdivided into westward and eastward moving types (WIG, EIG). Additionally, two special solutions exist: Mixed Rossby-gravity (MRG) waves behave as balanced modes when traveling westwards and as unbalanced waves when traveling eastwards. Kelvin waves have pure gravity wave properties in meridional direction and are balanced in the zonal direction. These eigenmodes have periods of less than a day to several weeks. Each wave type is associated with specific circulation patterns. Theoretical structures of all CCEWs are displayed in \citeauthor{Matsuno.1966} (1966, Figs.~4-8) and \citeauthor{Kiladis.2009} (2009, Fig.~3). 
	
	CCEWs also modulate the vertical thermodynamical structure of the tropical troposphere. Despite their different scales and circulation patterns, all CCEWs, except ER waves, have a self-similar, tilted vertical structure \citep{Mapes.2006,Kiladis.2009}. Moisture anomalies starting at low levels, rise to mid-levels prior to the convective peak. A stratiform moist outflow is left behind after the passage of the deep convection. Additional to adiabatic heating associated with the vertical circulation, diabatic affects create heating that slows down the wave. The reader is referred to \cite{Kiladis.2009} for a more detailed review of the theory, observational evidence, and properties of CCEWs.
	
	Two additional disturbances dominate rainfall variability over northern tropical Africa: The Madden-Julian Oscillation (MJO) is a wave mode that is not predicted by the shallow-water equations but shares several properties with Kelvin waves \citep{Madden.1971}. It is a global mode with highest activity over the Indian and Pacific Oceans. The MJO modulates the occurence of tropical rain clusters on the intraseasonal timescale and its influence has been documented for northern tropical Africa (e.g. \citealt{Matthews.2004,Pohl.2009,Alaka.2012}). Different theories have been proposed to explain its existence (see reviews by \citeauthor{Zhang.2005} 2005; \citeyear{Zhang.2013}). Tropical disturbances (TD) are westward travelling systems off the equator with a period of two to five days and are often precursors of tropical storms \citep{Riehl.1945,Roundy.2004}. Over northern tropical Africa, they mostly correspond to African Easterly Waves (AEWs), which a barotropic-baroclinic wave mode specific to the WAM system forming along the of the African Easterly Jet (AEJ, \citealt{Reed.1977,Duvel.1990}). They couple to convection and adominate variability at the daily timescale \citep{Fink.2003,Lavaysse.2006}. All CCEWs, the MJO, and TDs are collectively termed "tropical waves" in this paper.
	
	In the first part of this study, \citeauthor{Schlueter.2018} (2018, hereafter SFKV18) reviewed the modulation of rainfall by tropical waves. Depending on wave type, the modulation intensity varies from less than \SI{2}{\mm\per\day} to more than \SI{7}{\mm\per\day} during the monsoon season. Individual types can explain up to one third of rainfall variability on the daily to subseasonal timescales. The TD and Kelvin modes control precipitation on the daily timescales. ER waves and the MJO are dominate longer timescales (see SFKV18 and the references therein).
	
	Disturbances in the tropics are not independent from the extratropics. Using weather observations made during the Second World War, \citet{Riehl.1950} documented how tropical disturbances interact with the extratropical circulation. Predictability in the extratropics on the intraseasonal timescale largely originates from the tropics (\citealt{Waliser.2003, Vitart.2018}). The MJO is linked to several intraseasonal modes of variability, including the North Atlantic Oscillation (NAO) \citep{Walker.1932}, the Artic Oscillation (AO) \citep{Thompson.1998}, and the Pacific/North American Pattern (PNA) \citep{Wallace.1981}. A review of intreaseasonal tropical-extratropical interactions can be found in \citet{Stan.2017}.
	
	SFKV18 showed that the MJO and ER waves over northern tropical Africa are associated with rainfall far into the subtropics during the extended monsoon season. The eastward-tilted precipitation pattern resembles the tropical cloud plumes described by \citet{Knippertz.2003,Knippertz.2003b,Knippertz.2005} and \citet{Frohlich.2013}. Tropical plumes are more frequent during the winter half year when troughs triggered by breaking Rossby waves draw moist monsoonal air masses deep into the subtropics. During the monsoon season the subtropical jet (STJ) lies further to the north and thus tropical plumes are rare then \citep{Frohlich.2013}. It remains an open question whether the precipitation patterns associated with ER waves and the MJO are triggered by the same dynamical process as in wintertime tropical plumes.
	 
	The influence of tropical waves on rainfall depend on their influence on the local dynamics and thermodynamics. Several studies have analyzed the (thermo-) dynamic signatures of single wave types over this region in detail (e.g. TD: \citealt{Reed.1977,Duvel.1990, Fink.2003, Mekonnen.2006, Lavaysse.2006,Janiga.2016}, Kelvin: \citealt{Mounier.2007,Nguyen.2008,Mekonnen.2008,Ventrice.2013,Mekonnen.2016}, MJO: \citealt{Matthews.2004,Janicot.2009,Pohl.2009,Berhane.2015, Zaitchik.2017}, ER: \citealt{Janicot.2010, Yang.2018}, MRG: \citealt{Yang.2018,Cheng.2018}). The comparison of results for different waves is hampered by the use of different methodologies in the respective studies. So far no systematic comparison of all tropical waves has been performed using one consistent method. The aim of the present study is to close this gap and to provide a comprehensive analysis of the modulation of the WAM by tropical waves. This study will focus specifically on the effect of tropical waves on important components of the WAM such as the monsoon layer, inter-tropical discontinuity (ITD, \citealt{Lele.2010}) position, Saharan Heat Low (SHL, \citeauthor{Lavaysse.2009} 2009; \citeyear{Lavaysse.2010}) and AEJ. Additional attention is paid to the conditions for organized convection in form of mesoscale convective systems (MCSs, \citealt{Houze.2004}), which are responsible for the majority of rainfall in West Africa \citep{Eldridge.1957,Laing.1999, Fink.2003} and commonly cause extreme precipitation \citep{Engel.2017,Lafore.2017}. In order to make use of as much data as possible in this data scarce region, reanalysis data and in-situ measurements by radiosondes will be compared. More specifically the following research questions will be addressed:
	
	\begin{itemize}
		\item How do circulation patterns associated with tropical waves modify the WAM and the moisture distribution?
		\item How do tropical waves influence the vertical profile of temperature, moisture, and wind and what does this imply for the triggering and organization of rainfall?
		\item How is rainfall triggered over the Sahara by the MJO and ER waves and do they interact with the extratropical circulation?
	\end{itemize}
	
	In the following section the applied methods will be elaborated. The results will be presented in section \ref{sec:Results} and then discussed in section \ref{sec:Discussion}. The summary section in \ref{sec:Summary} concludes this study. 
	
	%%%%%%%%%%%%%%%%%%%%%%%%%%%%%%%%%%%%%%%%%%%%%%%%%%%%%%%%%%%%%%%%%%%%%
	% Methods
	%%%%%%%%%%%%%%%%%%%%%%%%%%%%%%%%%%%%%%%%%%%%%%%%%%%%%%%%%%%%%%%%%%%%%
	
	\section{Methods}  \label{sec:Methods}
	
	\subsection{Data}
	The modulation of dynamics and thermodynamics by tropical waves was examined using reanalysis data. Reanalysis products have the advantage of global coverage and a long temporal record. ERA-Interim is the most recent reanalysis product provided by the European Centre for Medium-Range Weather Forecasts (ECMWF) \citep{Dee.2011}, until it will be fully replaced by ERA-5. Several dynamical and thermodynamical fields were downloaded from the MARS archive (\url{http://apps.ecmwf.int/datasets/}) and analyzed in this study. 
	
	The daily interpolated National Oceanic and Atmospheric Administration (NOAA) OLR dataset \citep{Liebmann.1996} was used to filter the waves following SFKV18. OLR has the advantage that it is available for a relatively long period. Consistent with SFKV18, the analyzed time period spans from 1981 to 2013. The precipitation patterns associated with each wave band are discussed in detail in SFKV18,. Following SFKV18, the Climate Hazards Group InfraRed Precipitation with Station data V.2 (CHIRPS) dataset was used to visualize the associated rainfall anomaly patterns. CHIRPS estimates daily precipitation rates over land masses only by gauge-calibration of infra-red measurements and has a spatial resolution of \ang{0.5}$\times$\ang{0.5} \citep{Funk.2015}. 
	
	Reanalyses have large biases in humidity fields but also in wind and temperature over the measurement-sparse African continent \citep{AgustiPanareda.2010,AgustiPanareda.2010b,Roberts.2015}. Thus, in-situ measurements by radiosondes are a valuable supplement to accurately assess the modulation of atmospheric profiles. For this purpose, quality-controlled radiosonde measurements from the Integrated Global Radiosonde Archive (IGRA) Version 2 \citep{Durre.2016} were analyzed, which were downloaded from \url{https://data.nodc.noaa.gov/cgi-bin/iso?id=gov.noaa.ncdc:C00975}. Due to a relatively high data availability, 12 UTC ascents at Abidjan Airport (\ang{13.48}N, \ang{2.17}E) and Niamey Airport (\ang{5.25}N, \ang{3.93}W) were chosen to illustrate the modulation over the Guinea Coast box (\ang{5}W-\ang{5}E, \ang{5}-\ang{10}N) and the West Sahel box (\ang{5}W-\ang{5}E, \ang{10}-\ang{15}N) defined in SFKV18. Composites of radiosonde ascents from Abidjan were calculated for the more equatorial MRG and Kelvin waves, as both are predominantly equatorial phenomena, whereas the influence of ER waves and TDs was analyzed using the Niamey data, as both wave modes have a stronger imprint farther to the north (see Figs.~4-6 in SFKV18). Between April to October 1981 to 2013, 1370 measurements are available for Niamey (\SI{19.4}{\percent} of the study period) and 1214 measurements (\SI{17.2}{\percent}) for Abidjan. As reference for the observed rainfall, the daily rain gauge observations were obtained from the Karls\-ruhe African Surface Station Database (KASS-D). 
	
	The influence of the MJO was compared with the NAO, AO, and PNA. Daily normalized indices for these modes were downloaded from \url{ftp://ftp.cpc.ncep.noaa.gov/cwlinks}. Further information on how the indices are calculated can be found at \url{http://www.cpc.ncep.noaa.gov/products/precip/CWlink/daily_ao_index/history/method.shtml}.
	
	\subsection{Wave filtering}
	The same approach as in SFKV18 was applied to identify tropical waves. In a nutshell, OLR was filtered using the classical method by \cite{Wheeler.1999}. The same filtering bands for the six waves MJO, ER, MRG,  Kelvin, TD, and EIG were applied (Tab.~1 in SFKV18). No symmetry was imposed. Background noise also projects into the filter bands \citep{Wheeler.1999}, but this is likely removed when compositing over many cases as done here. It should be noted that filtered signals are related to convectively coupled waves and not to dry equatorial waves because low OLR indicates convective activity. Thus, conclusions can not be readily transferred to the impact of dry equatorial waves on the WAM system.
	
	\subsection{Composite study}
	A local wave phase metric is constructed based on the wave signal at \ang{0}E and its time derivative \citep{Riley.2011, Yasunaga.2012}. The phase-amplitude-space is divided into eight phases, excluding days when the amplitude is not significant. The reader is referred to SFKV18 for further details of the method.
	
	Composites of several dynamic and thermodynamic fields were calculated for all days when the respective wave is in each phase. The moisture transport is analyzed using anomalies of moisture flux, moisture flux convergence, precipitable water (PW), and CHIRPS precipitation during all phases (Figs.~\ref{fig.mapMJO}--\ref{fig.mapTD}). The circulation patterns as seen from wind, geopotential, and divergence fields are consistent and thus only provided in the supplementary material (SM) (Figs.~S1--S5). The position of the ITD gives an estimate of the northward extent of the monsoon layer. The ITD is commonly indicated by the isoline of \SI{14}{\celsius} dew point temperature at \SI{2}{\meter} \citep{Buckle.1996}. Plots of wind, and geopotential in \SI{850}{\hecto\pascal}, as well as divergence in \SI{200}{\hecto\pascal} are provided in the SM. Geopotential in \SI{300}{\hecto\pascal} is shown for the MJO and ER waves instead of \SI{850}{\hecto\pascal} to highlight their relationship with the extratropical wave guide.
	
	The influence of tropical waves on the organization of deep convection was analyzed using three variables that are known to be key ingredients for MCSs \citep{Maranan.2018}: Convectively available potential energy (CAPE) is a general measure for atmospheric instability  and the strength of convective updrafts \citep{Moncrieff.1976}. Secondly, dry mid-levels intensify evaporative downdrafts in MCSs \citep{Raymond.1990, Brown.1997}. These downdrafts induce cold pools, which reinforce the MCS \citep{Zipser.1977,Corfidi.2003}. The strength of downdrafts was estimated using relative humidity in \SI{500}{\hecto\pascal} (RH\textsubscript{500}) following \cite{Roca.2005}. Finally, low-level wind shear is necessary for MCSs to separate up- and downdrafts and support longevity of the system \citep{Browning.1962,Rotunno.1988}. Here, we use the total wind difference between 600 and \SI{925}{\hecto\pascal} (Shear\textsubscript{600-925}). Mean CAPE, RH\textsubscript{500}, and Shear\textsubscript{600-925} anomalies were calculated for each phase in the Guinea Coast and West Sahel boxes.
	
	Mean anomalies of all fields were calculated with respect to the climatology from 1979 to 2016. Significance was tested at the \SI{5}{\percent}-level using non-parametric bootstrapping and $n=1000$ repetitions. Wind anomalies are considered significant when either the zonal or meridional anomaly was significantly different from zero. A \ang{5}-spatial smoothing was applied to divergence, moisture flux divergence, and geopotential fields after compositing.
	
	\subsection{Radiosonde study}
	\label{sec:methods_radiosonde}
	Most radiosonde ascents lack measurements above \SI{300}{\hecto\pascal}, and thus only RH below \SI{300}{\hecto\pascal} was analyzed. The raw data were read with the python-package \textit{IGRA2reader} and interpolated to every \SI{5}{\hecto\pascal} and then composites of all ascents during all eight wave phases were constructed. In order to capture the local wave signal over the station, the wave phases were determined for a narrower Guinean and Sahelian zonal wave band. In Abidjan, wave phases were calculated based on \ang{5}--\ang{10}N, \ang{4}W, in Niamey based on \ang{10}--\ang{15}N, \ang{2}E. The absolute difference of the wind vector at 600 and \SI{925}{\hecto\pascal} was calculated for each profile and averaged for each phase. 
	
	To analyze the thermodynamic conditions during the passage of the waves, mixed layer parcel buoyancy of the mean profile during all phases was calculated following the approach in \citet{Schrage.2006}. The parcel buoyancy at a pressure level with the environmental temperature $T_e$ is defined as 
	
	\[ B = T_p - T_e,\] 
	
	where $T_p$ is the parcel profile ascending dry adiabatically until the lifting condensation level (LCL) and moist adiabatically above the LCL. The parcel profile was determined based on the mean temperature and humidity of the mixed layer between 15 and \SI{65}{\hecto\pascal} above ground (approximately \SI{400}{\meter}). The lowest \SI{10}{\hecto\pascal} were removed before the parcel profile was calculated due to contamination from inconsistent surface measurements that stem from radiosonde initialization. Parcel buoyancy can be used to analyze the vertical stability. By definition $B=0$ at the LCL and at the equilibrium level (EL). The integral of $B$ between the surface and LCL is proportional to CIN, and the area between the LCL and EL to CAPE.
	
	The rainfall signal was plotted using rain gauge data from the stations, where the radiosondes were launched. Rain gauges report daily rainfall from 06 to 06 UTC+1d, whereas radiosonde are launched at 12 UTC. The shift of six hours was eliminated using the method as presented in SFKV18: The mean periods of the respective waves were calculated using MATLAB-function \textit{meanfreq()}. Based on this, the length of six hours measured in wave phases was estimated in order to shift the precipitation curve to match the radiosonde data.	
	
	\subsection{Time-lagged analysis}
	The MJO and ER waves trigger rainfall anomalies up to the Mediterranean Sea (SFKV18). To test, whether and how the waves couple with the extratropical circulation, geopotential and wind in \SI{300}{\hecto\pascal} were analyzed. The influence on lower-tropospheric thickness was analyzed between 600 and \SI{925}{\hecto\pascal}. The origin and development of the wave signal were traced using a time-lag analysis. The tropical plumes were observed in the local phase 4 (see SFKV18, section 2~d for more detail on the phase diagram) for the MJO and in phase 6 for ER waves. Composites of precipitation, geopotential, and wind in \SI{300}{\hecto\pascal} were calculated for 25 days to 5 days after the MJO wave was in phase 4 and 15 to 3 days after the ER was in phase 6.
	
	%%%%%%%%%%%%%%%%%%%%%%%%%%%%%%%%%%%%%%%%%%%%%%%%%%%%%%%%%%%%%%%%%%%%%
	% Results
	%%%%%%%%%%%%%%%%%%%%%%%%%%%%%%%%%%%%%%%%%%%%%%%%%%%%%%%%%%%%%%%%%%%%%
	
	\section{Results}  \label{sec:Results}
	
	This section will describe first the circulation patterns of the different wave types and their influence on the moisture distribution. Then, their vertical structure will be examined and differences in mechanisms of rainfall modulation will be shown. The last part of this section will analyze how rainfall over the Sahara is triggered by the MJO and ER waves and whether they interact with the extratropical circulation.
	
	\subsection{Circulation patterns and moisture modulation}

	\begin{figure*}[p]
		\centering
		\noindent\includegraphics[height = 0.8 \textheight ]{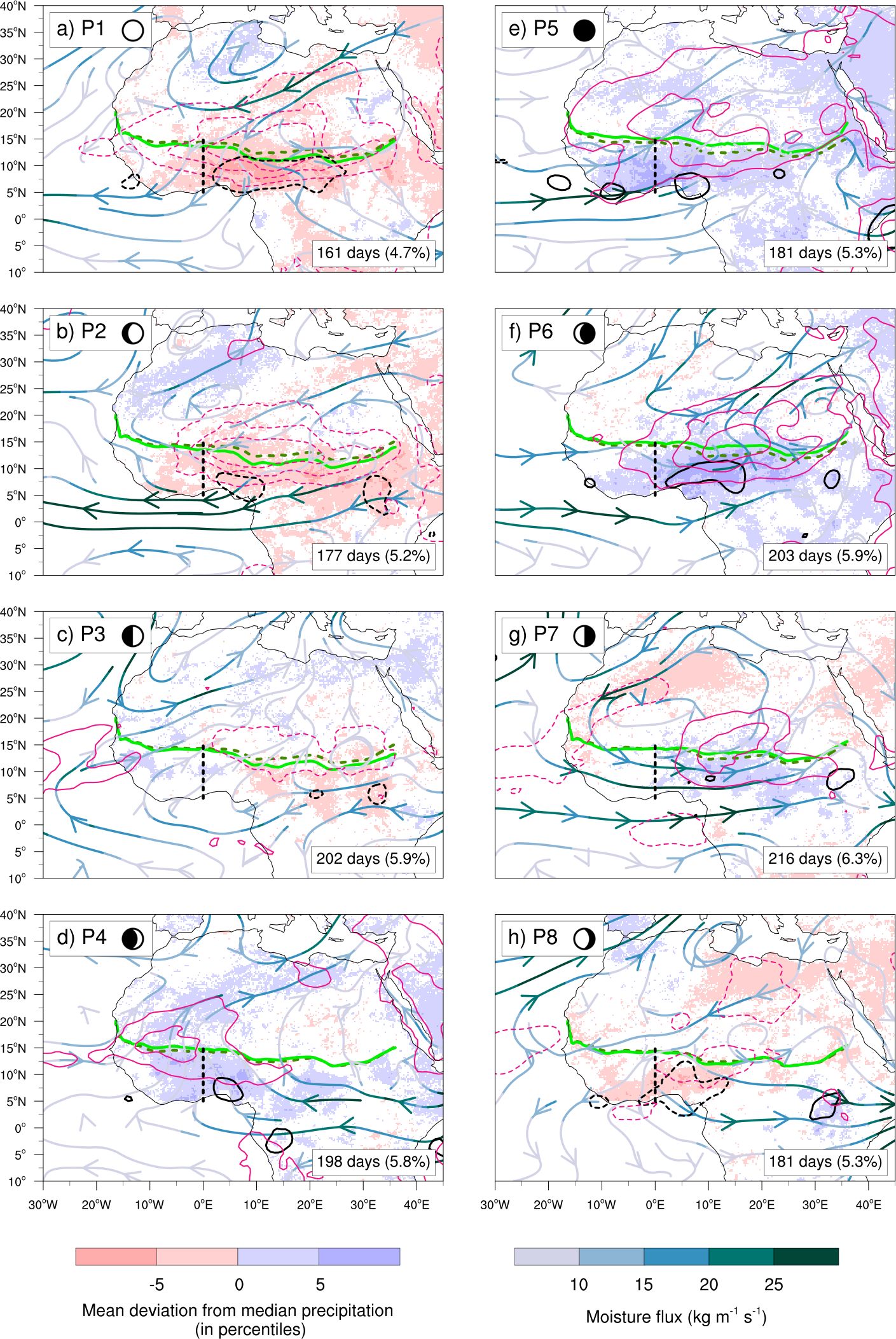}\\
		\caption{Composite of moisture flux (streamlines, magnitude in turquoise colors), moisture flux convergence (black lines, at $\pm$ 1 and \SI{2e-5}{\kg\per\meter\squared\per\second}, negative dashed), and PW (pink lines, at $\pm$ 1, 2 and \SI{3}{\mm}, negative dashed) for days with significant Madden-Julian Oscillation signals over \ang{5}-\ang{15}N, \ang{0}E (dashed vertical line) during the extended monsoon season (March--October). Significant precipitation anomalies are shaded (see SFKV18, their Figs. 5 and 6). The mean position of the ITD in each phase is depicted with a green line. The dashed dark green line shows the mean position in all phases. The number of days and the fraction of days per season during each phase is indicated in the bottom right corner of each panel. Small moon symbols indicate the phase of the wave, with dry anomalies in white and wet anomalies in black.}
		\label{fig.mapMJO}
	\end{figure*}
	
	The MJO strongly modulates the circulation over entire African (Figs.~\ref{fig.mapMJO} and S1). Consistent with theory, the wet phase over northern tropical Africa lies in a region of enhanced westerlies, which lead to increased cyclonic shear vorticity, and the dry phase in a region of easterlies. Low-level convergence and upper-level divergence are associated with lifting of air parcels and, thus, increased precipitation (Fig.~S1). Additional to the modulation of zonal winds, vortex-like structures exist, which are located at \ang{20}--\ang{25}N (e.g., Figs.~\ref{fig.mapMJO}b,f,g and S1c,d,g). \SI{300}{\hecto\pascal} geopotential composites indicate a relationship with extratropical Rossby waves (Fig.~S1), which will be discussed in section \ref{sec:Rossby}. The MJO has a large-scale influence on PW. The monsoonal flow intensifies during wet phases and reduces during dry phases. In central and eastern Africa, the ITD moves southward by 100-\SI{200}{\kilo\meter} during the dry phases and northward during the wet phases. West of \ang{5}E, the modulation of the ITD is weaker. The SHL is ventilated by increased monsoonal and Atlantic air masses. Significant moisture anomalies can be found far into the subtropics (>\ang{30}N). Due to the low frequency of the MJO, the total moisture convergence is small compared to the other wave types.
	
	\begin{figure*}[p]
		\centering
		\noindent\includegraphics[height = 0.95 \textheight ]{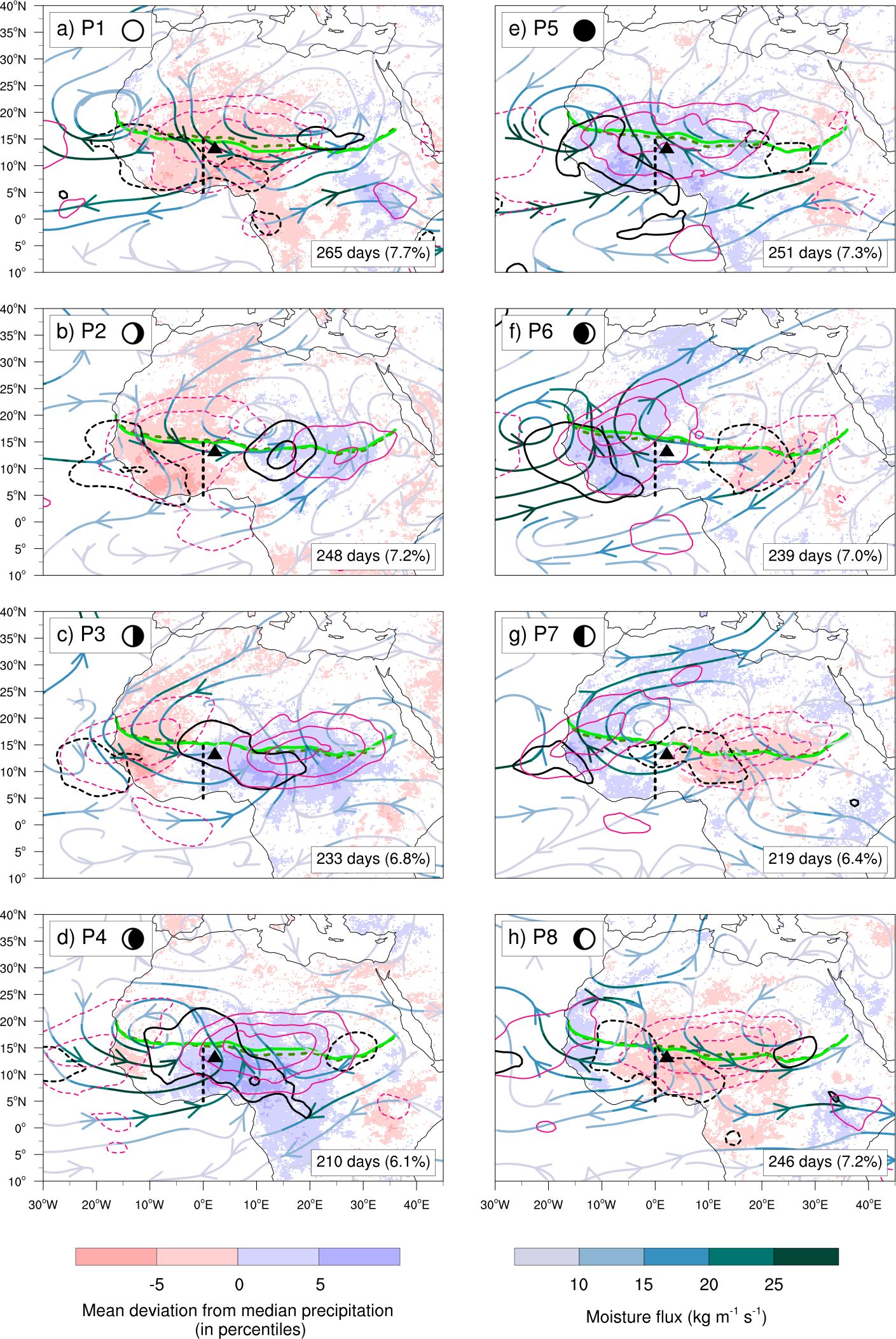}\\
		\caption{Same as Fig.~\ref{fig.mapMJO}, but for equatorial Rossby wave. The triangle shows the location of the radiosonde station in Niamey used in Fig.~\ref{fig.radiosondes}a.}\label{fig.mapER}
	\end{figure*}
	
	The more rotational ER waves show westward propagating vortices with centers off the equator at around \ang{20}N (Figs.~\ref{fig.mapER} and S2). Additional to the vortices in the tropics, ER waves are associated with upper-level troughs and ridges over the subtropical Atlantic Ocean and Mediterranean (Fig.~S2). The link to extratropical Rossby waves will be discussed in section \ref{sec:Rossby}. Sustained moderate moisture convergence during the passage of the slow-moving ER waves result in a strong modulation of PW and precipitation is mainly aligned with these anomalies. Moist air masses from the Gulf of Guinea and Congo Basin are transported towards tropical West and Central Africa during the wet phase. In the Sahel, wet conditions are related to enhanced monsoonal flow and dry conditions with decreased monsoonal inflow. The mean position of the ITD is roughly \SI{100}{\kilo\meter} further north in the wet sector of the wave and further south in the dry sector. The SHL is ventilated by southerlies during phases 5--7. These southerlies result in moisture and precipitation anomalies reaching up to the Mediterranean (Fig.~S2).
	
	\begin{figure*}[p]
		\centering
		\noindent\includegraphics[height = 0.95 \textheight ]{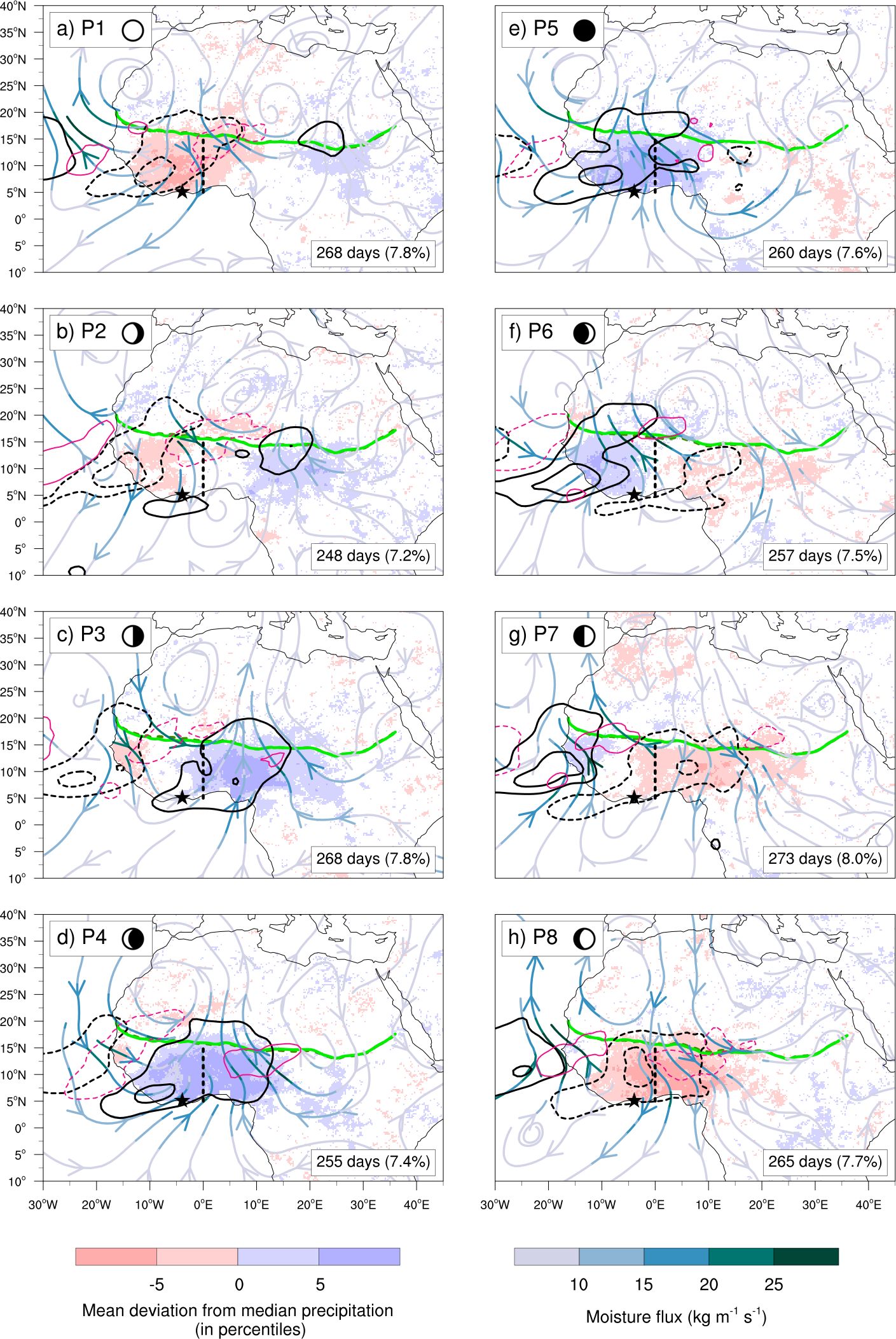}\\
		\caption{Same as Fig.~\ref{fig.mapMJO}, but for mixed Rossby gravity wave. The star shows the location of the radiosonde station in Abidjan used in Fig.~\ref{fig.radiosondes}b.}\label{fig.mapMRG}
	\end{figure*}
	
	MRG waves are also a more rotational wave type. They feature deep westward moving vortices reaching from 850--\SI{300}{\hecto\pascal} and centered at around 25-\ang{30}N (Figs.~\ref{fig.mapMRG} and S3). Consistent with theory, they modulate the cross-equatorial meridional flow. Precipitation is mainly related to moisture flux convergence. Similar to the faster moving Kelvin waves and TDs, the modulation of PW is weaker compared with the MJO and ER waves. Moist air masses are then transported northward originating from the Gulf of Guinea and Congo Basin. These moisture anomalies reach up to \ang{20}N. The mean ITD position, however, is not significantly affected. Moisture flux convergence directly at the Guinea Coast leads the anomalies farther inland. In the subtropics, a westward propagating signal can be observed at \SI{300}{\hecto\pascal} (Fig.~S3). 
	
	\begin{figure*}[p]
		\centering
		\noindent\includegraphics[height = 0.95 \textheight ]{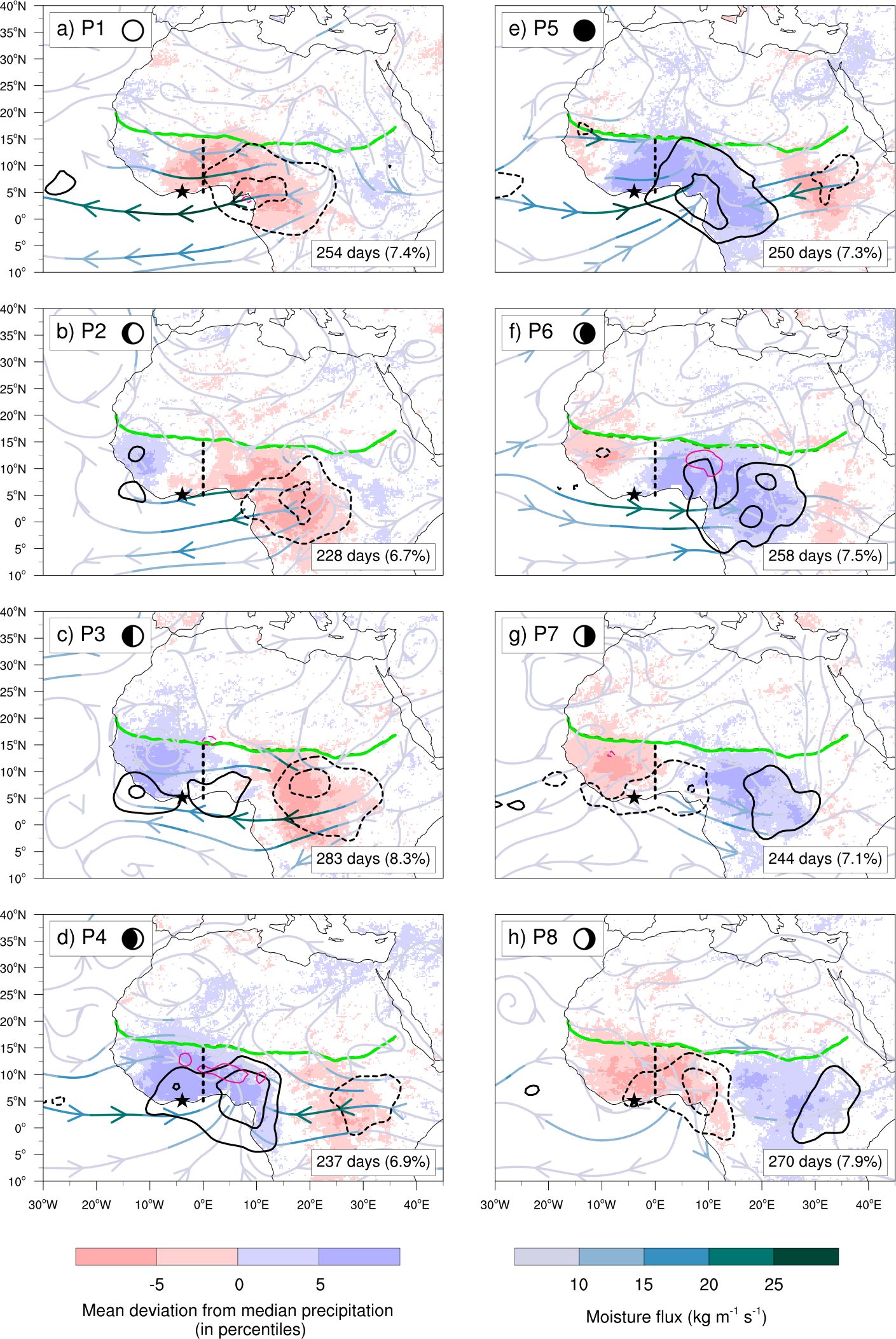}\\
		\caption{Same as Fig.~\ref{fig.mapMJO}, but for Kelvin wave. The star shows the location of the radiosonde station in Abidjan used in Fig.~\ref{fig.radiosondes}c.}\label{fig.mapKelvin}
	\end{figure*}
	
	Kelvin waves show a pronounced influence on the zonal flow patterns (Figs.~\ref{fig.mapKelvin} and S4). As predicted by theory, wet phases are associated with regions of enhanced westerlies and vice versa. Kelvin waves are vertically tilted; the lower-level convergence center leads the precipitation anomalies, with upper-level divergence lagging by approximately $1/8$ of the wavelength (Fig.~S4). Although the flow pattern is primarily zonal, an enhanced southerly component is evident in the wet phase and a northerly component in the dry phase over West Africa. Similar to the MJO, westerlies during the wet phase lead to increased horizontal shear on the southern flank of the AEJ. Kelvin waves facilitate precipitation mainly due to moisture flux convergence. Because moisture has a low gradient meridionally (not shown), the moisture flux convergence is predominantly generated due to convergence of zonal winds. The mean ITD position does not shift significantly. The influence on precipitation and circulation reaches into the Sahel to about \ang{15}N. As Kelvin waves quickly move over Africa and moisture is removed by precipitation, PW is not strongly affected.
	
	\begin{figure*}[p]
		\centering
		\noindent\includegraphics[height = 0.95 \textheight ]{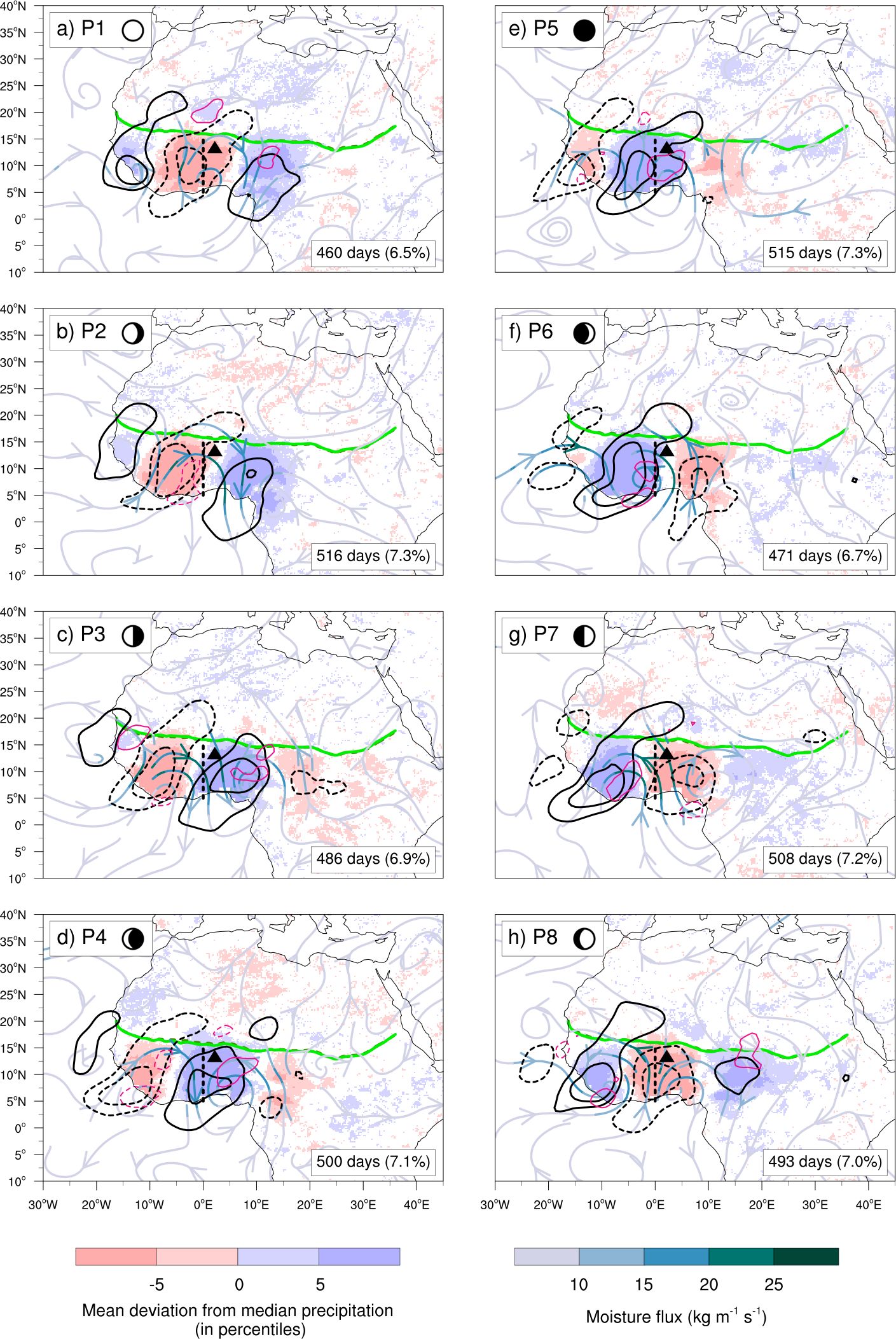}\\
		\caption{Same as Fig.~\ref{fig.mapMJO}, but for tropical depression (African Easterly Wave). The triangle shows the location of the radiosonde station in Niamey used in Fig.~\ref{fig.radiosondes}a.}\label{fig.mapTD}
	\end{figure*}
	
	TDs have a strong influence on the upper-level divergence field (Figs.~\ref{fig.mapTD} and S5) due to their strong coupling with convection. The flow is modulated up to \ang{20}--\ang{25}N. As documented in previous studies, enhanced convection can be found in the trough of the AEW and ahead of it, whereas suppressed convection is evident in and ahead of the ridge. A weak cross-equatorial flow might stem from contamination by MRG waves, which have a partial overlap in the filter bands with TDs (see SFKV18). The vortices related to the modulation of the AEJ, result in moisture flux convergence and divergence (Fig.~\ref{fig.mapTD}). Additionally, low-level divergent outflow in the south-western and north-eastern parts of the vortices lead to moisture flux divergence, whereas convergent inflow in these regions leads to moisture flux convergence. The resulting tilt can also be seen in TRMM precipitation (Fig.~S11 in SFKV18). In the wake of the precipitation maxima and minima, PW is slightly reduced and enhanced, respectively. Associated with the northerlies in the dry sector and southerlies in the wet sector, the mean position of the ITD modulates by about 50 to \SI{100}{\kilo\meter}.
	
	The influence of EIG waves on the circulation and moisture fields is weak and, thus, their plots are only shown in the SM (Figs.~S6 and S7). Regions of easterlies are associated with upper-level divergence and positive precipitation anomalies, and vice versa. The moisture distribution and ITD position are hardly influenced.
	
	\subsection{Vertical structure and mechanism of rainfall modulation}
	\begin{figure*}
		\centering
		\noindent\includegraphics[height = 0.92 \textheight ]{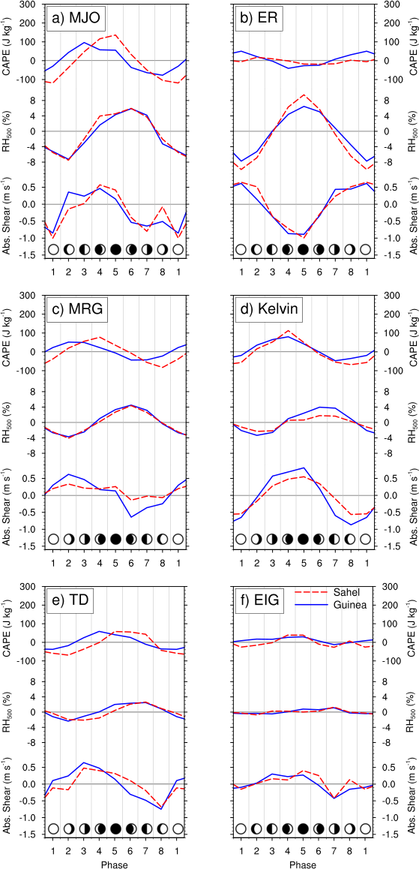}\\
		\caption{Modulation of CAPE, relative humdity in \SI{500}{\hecto\pascal} (RH\textsubscript{500}), and the low-level wind shear between 600 and \SI{925}{\hecto\pascal} in the Sahel box (dashed red line) and Guinea box (solid blue line) during all 8 phases for six tropical waves (a-f). Small moon symbols indicate the phase of the wave, with dry anomalies in white and wet anomalies in black (see Figs.~\ref{fig.mapMJO}--\ref{fig.mapTD})}\label{fig.lines}
	\end{figure*}
\begin{figure*}
		\centering
		\noindent\includegraphics[width = 13.8cm]{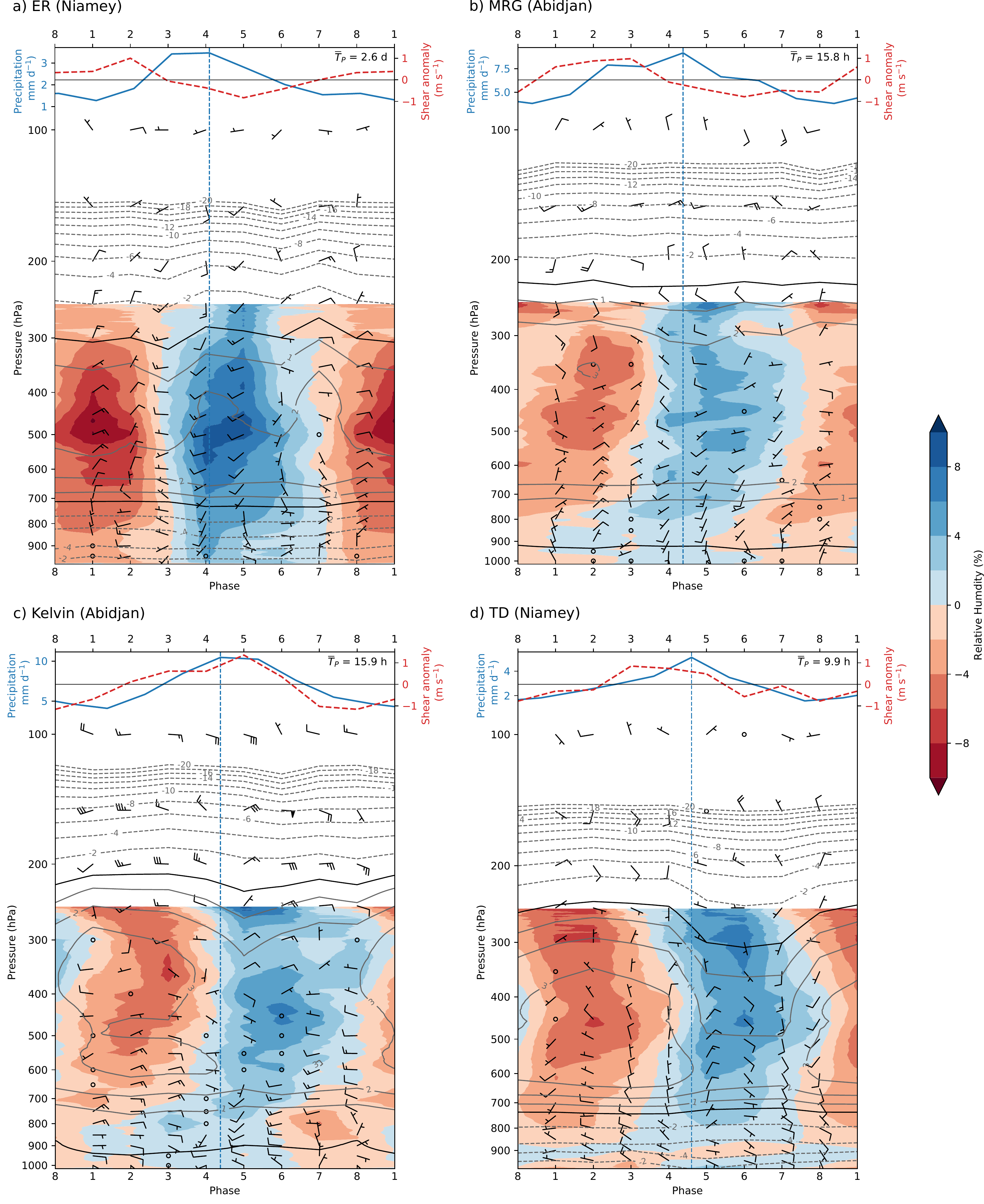}\\
		\caption{Radiosonde composite for all phases of (a) ER waves over Niamey, (b) MRG waves over Abidjan, (c) Kelvin waves over Abidjan, and (d) TDs over Niamey during April to October. Filtering was done at \ang{2}E and \ang{10}--\ang{15}N for Niamey \ang{2}E and \ang{5}--\ang{10}N for Abidjan. Top of panels shows absolute wind shear anomaly between 600 and \SI{925}{\hecto\pascal} (red dashed line) and precipitation anomaly (blue line) as recorded by the rain gauge at the respective location. The precipitation is shifted by 6h to account for the time difference between radiosonde and rain gauge measurement (converted by average phase length $\overline{T}_P$, see method section \ref{sec:Methods}\ref{sec:methods_radiosonde} for more detail). The blue dashed vertical line indicates peak of rainfall. Lower parts of panels show relative humidity anomaly (shading), parcel buoyancy anomaly (contours, negative dashed), and horizontal wind anomaly (wind barbs, half barbs equal to \SI{0.5}{\meter\per\second}, full barbs to \SI{1}{\meter\per\second}, and triangles to \SI{5}{\meter\per\second}). Bottom thick black contour line represents the LFC, and top thick line the EL.}\label{fig.radiosondes}
	\end{figure*}

	The mechanisms of rainfall modulation depend on the vertical profile of the tropical waves. Figure~\ref{fig.lines} shows the modulation of three key ingredients for convective organization in two boxes in West Africa as measured by reanalysis data. To give a more detailed picture of the vertical structure of ER, Kelvin, and MRG waves, as well as TDs, radiosonde data are also analyzed (Fig.~\ref{fig.radiosondes}). No consistent modulation of radiosonde data was found for the MJO and EIG wave and, thus, their composites are not shown here.
	
	The MJO facilitates organization of deep convection. CAPE in the Guinea Coast leads the convecti maximum by two phases and is in phase with convection in the West Sahel (Fig.~\ref{fig.lines}a). Before the wet peak, absolute wind shear increases and decreases sharply after the peak. Convective down-drafts are likely weaker during wet phases, as mid-level moisture is enhanced. Conditions for organized convection are favorable before and during the wet peak of rainfall and slightly unfavorable after the peak in rainfall.
	
	ER waves (Fig.~\ref{fig.lines}b) show a unique modulation of thermodynamic conditions with unfavorable conditions for MCSs during the wet phases. In contrast to the other wave types, CAPE and wind shear are reduced during wet phases and enhanced during dry phases, although the CAPE anomaly is weak. Mid-level moisture is strongly enhanced during the wet phases of ER waves and decreased during dry phases. Radiosonde data at Niamey gives a more detailed view on the vertical structure of ER waves (Fig.~\ref{fig.radiosondes}a). In contrast to the following waves, ER waves, are not tilted vertically. During the passage of ER waves, the entire troposphere moistens and then dries. In the mid-troposphere, absolute RH increases and decreases by more than \SI{10}{\percent}. Consistent with reanalysis data, the effect on parcel buoyancy is weak. LCL and CIN are slightly lowered during the wet phase. Winds turn cyclonically with time as expected from Fig.~\ref{fig.mapER}. The AEJ is weaker before the wet phase and stronger afterwards. Upper-level wind divergence lead the rainfall maximum. This indicates that precipitation is mainly generated by large-scale lifting and moistening.
	
	The influence of MRG waves (Fig.~\ref{fig.lines}c) on the thermodynamic environment is in between the MJO and ER. CAPE anomalies lead precipitation anomalies by three phases in the Guinea Coast and by one phase in the West Sahel. Mid-level moisture is enhanced after the peak of the wet phase and suppressed after the dry phase. Wind shear anomalies lead precipitation anomalies by about two phases with a weaker modulation in the West Sahel. Radiosondes ascents at Abidjan reveal a vertical tilt. (Fig.~\ref{fig.radiosondes}b). Moisture builds up at low levels (900--\SI{800}{\hecto\pascal}) two to three phases before the peak in rainfall. The moisture anomaly shifts upward and, eventually, the entire troposphere is moistened during the rainfall peak. Upper- and mid-level moisture anomalies persist after the wet peak, while dry anomalies build up at lower levels (phases 5--7). MRG waves only weakly modulate parcel buoyancy. CAPE is enhanced before the build-up of deep convection and reduces during it. Low-level winds converge during the wet peak. The AEJ is stronger during the dry phase and weakened during the wet phase. Upper-level winds converge before the wet peak and diverge after the wet peak, indicating that deep convection is first suppressed but emerges during the wet peak. Low-level wind shear is enhanced before the wet peak and reduces after the rainfall peak. Therefore, good conditions for organized convection persist only during the first part of the wet phases. The build-up of rainfall by MRG waves is gradual over Abidjan until deep convection develops followed by a stratiform outflow region (Fig.~\ref{fig.radiosondes}b). 
	
	Kelvin waves (Fig.~\ref{fig.lines}d) have similar modulation signature as the MJO but on a shorter timescale. CAPE leads precipitation by about one phase. Mid-level moisture is reduced prior to the peak of rainfall and enhanced after the peak of rainfall. As Kelvin waves modulate zonal winds (Figs.~\ref{fig.mapKelvin} and S4), low-level wind shear is strongly modulated, which leads the wet peak by half a phase. The radiosonde measurements at Abidjan reveal a vertical tilt (Fig.~\ref{fig.radiosondes}c). Moisture anomalies build up at 900--\SI{800}{\hecto\pascal} three phases before the peak in rainfall, rise, and finally the entire troposphere is moistened after the peak of rainfall. Upper- and mid-level moisture anomalies persist after the wet peak, while lower levels start to dry again (phases 7--8). Parcel buoyancy is strongly modulated. CAPE is enhanced before the peak of rainfall and reduces with the onset of rain. During the deep convection, CIN increases. Then, the LCL is more than \SI{50}{\hecto\pascal} higher than before the peak rainfall and the EL is lowered by about \SI{20}{\hecto\pascal}. The vertical tilt of the Kelvin wave is also evident in zonal wind anomalies, which move upward with time. Meridional winds are only weakly modulated. Low-level wind shear is enhanced during the wet phase. The AEJ is strengthened during and after the dry phase, and reduces after the peak of rainfall. Upper-level winds are strongly modulated. The tropical easterly jet is enhanced during phases 5-7 and weakened during the phases 8--2. Upper-level winds diverge during the peak of rainfall facilitating deep convection. In summary, the radiosonde data suggest a build-up of low-level clouds before the rainfall peak, followed by deep and well organized convection, and finally a stratiform outflow region.
	
	The influence of TDs (Fig.~\ref{fig.lines}e) on organization of convection resembles the modulation by Kelvin waves. The phasing and magnitude of CAPE, mid-level moisture, and wind shear anomalies are remarkably similar. Wind shear anomalies, however, are slightly weaker and do not change as abruptly compared to Kelvin waves. TDs are also tilted vertically, as the profile over Niamey shows (Fig.~\ref{fig.radiosondes}d). A very shallow region of positive humidity anomalies exists at \SI{900}{\hecto\pascal} during the dry phase. During phases 3--4, the troposphere moistens relatively rapidly, starting from the lower and the upper troposphere. By phase 5, the entire troposphere is moistened. Moist anomalies remain at 500--\SI{350}{\hecto\pascal} after the passage of rain. Parcel buoyancy is strongly modulated with CAPE notably enhanced and CIN reduced before the wet peak, which is quickly diminished and increased, respectively, when deep convection occurs. After the peak of rainfall, the LCL is lifted and the EL is lowered. Consistent westerly (easterly) anomalies are evident in the lower troposphere during the dry (wet) phase. The AEJ is modulated and turns anti-cyclonically with time, whereas upper-level winds turn cyclonically. Wind shear increases before and during the rainfall peak. It can be concluded that rainfall is embedded in a circulation system that facilitates the formation of MCSs.
		
	EIG waves (Fig.~\ref{fig.lines}f) only weakly modulate CAPE, mid-level moisture, and wind shear. Wind shear is weakly enhanced before, and reduced after the wet phase. Organized convection is not likely affected by the passage of EIG waves.
	
	\subsection{Relationship to extra-tropical Rossby waves}
	\label{sec:Rossby}
	\begin{figure}
		\centering
		\noindent\includegraphics[clip, trim=7cm 7.5cm 7cm 7.5cm, width=5cm]{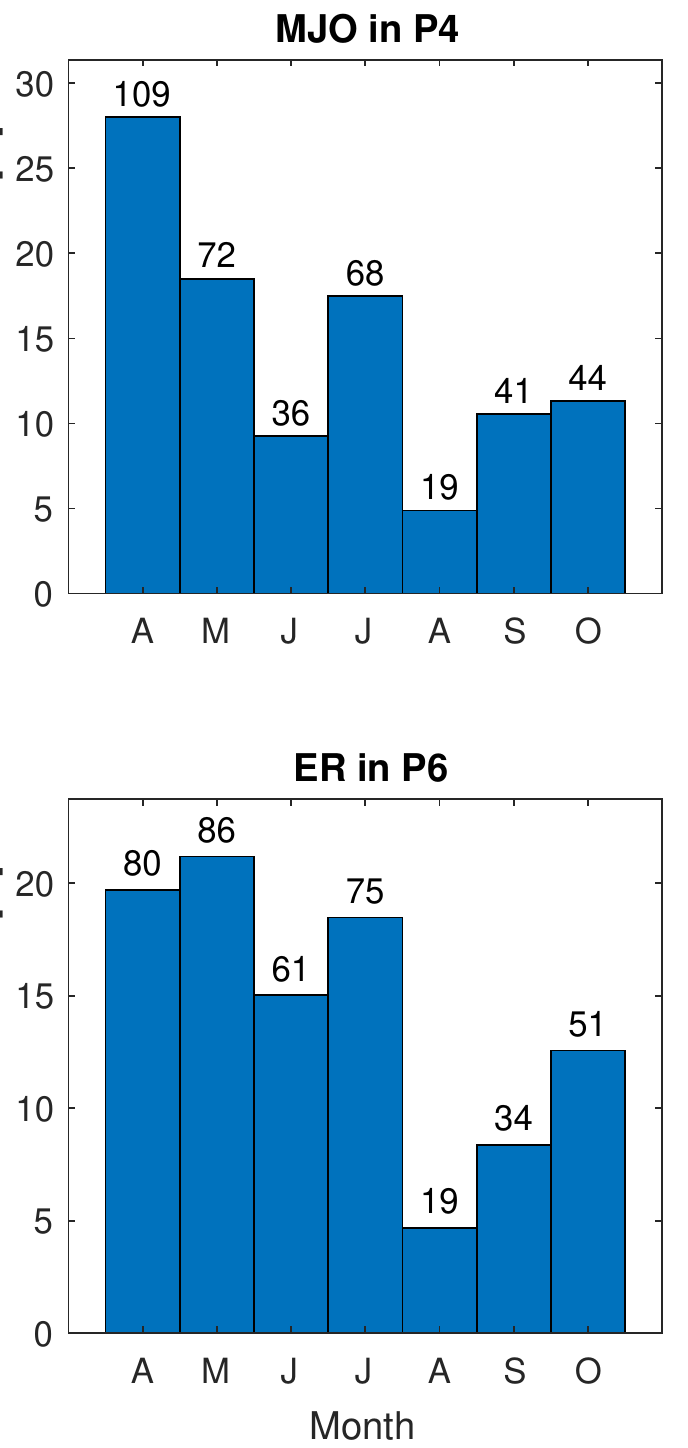}\\
		\caption{Relative occurrence of months during which local (a) MJO is significant in phase 4 and (b) ER in phase 6 at \ang{0}E. Numbers over the bars show total count of days during each month.}
		\label{fig.months_MJO_ER}
	\end{figure}

\begin{figure}
		\centering
		\noindent\includegraphics[clip, trim=7cm 10cm 7cm 10cm, width=5cm]{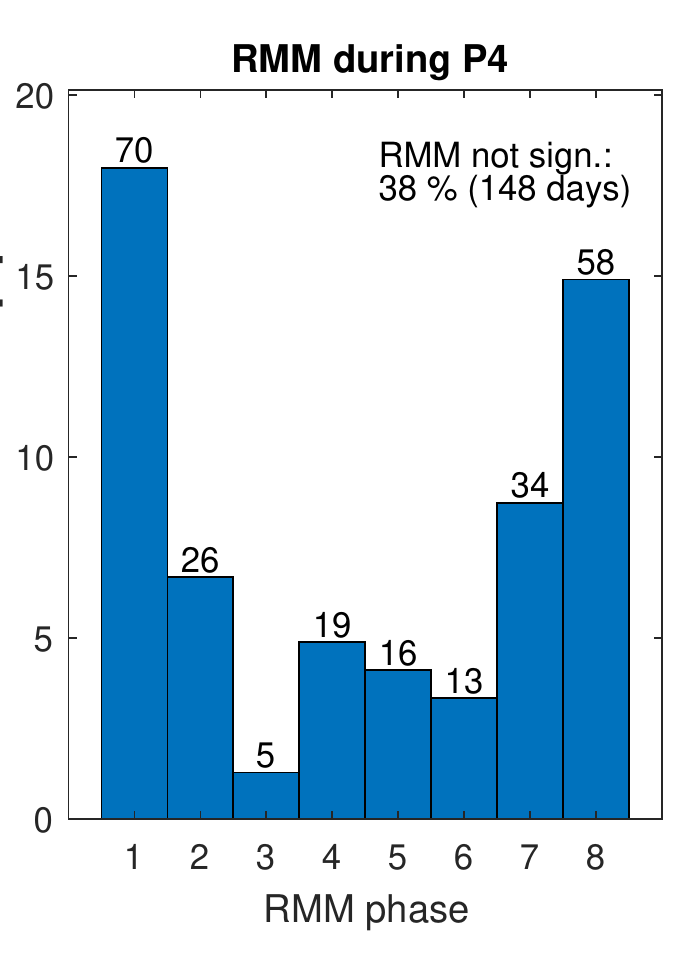}\\
		\caption{Real-time multivariate MJO index (RMM) when local MJO is significant in phase 4. The histogram depicts the relative occurrence in each phase. Numbers over the bars show total count of days during each RMM phase. The index was obtained from http://www.bom.gov.au/climate/mjo/.}
		\label{fig.RMM_P4}
	\end{figure}

	The MJO and ER waves can trigger tropical plume-like precipitation anomalies over the Sahara. These structures appear during the local phase 4 of the MJO and phase 6 of ER. Both wave modes are more active before the peak of the WAM. Between April and July, \SI{73}{\percent} of cases with MJO in phase 4 occur and \SI{74}{\percent} of cases with ER in phase 6 (Fig.~\ref{fig.months_MJO_ER}). The preference of tropical plumes during spring is even more evident when only looking at strong cases in phase 4 and 6, respectively, where the local wave amplitude is greater than 2. \SI{90}{\percent} of the 97 strong MJO cases during the extended monsoon season and \SI{76}{\percent} of the 106 strong ER cases take place during April and May (not shown). This study uses a local wave filtering. This way the local "flavor" of the MJO is captured more adequately. To allow comparison with the widely used real-time multivariate MJO index (RMM, \citealt{Wheeler.2004}), Fig.~\ref{fig.RMM_P4} shows the RMM during the local phase 4, when the plume is observed. The MJO as a global mode is over the western hemisphere and Africa (RMM Phases 1 and 8).  
	
\begin{figure*}
		\centering
		\noindent\includegraphics[height = 0.89 \textheight]{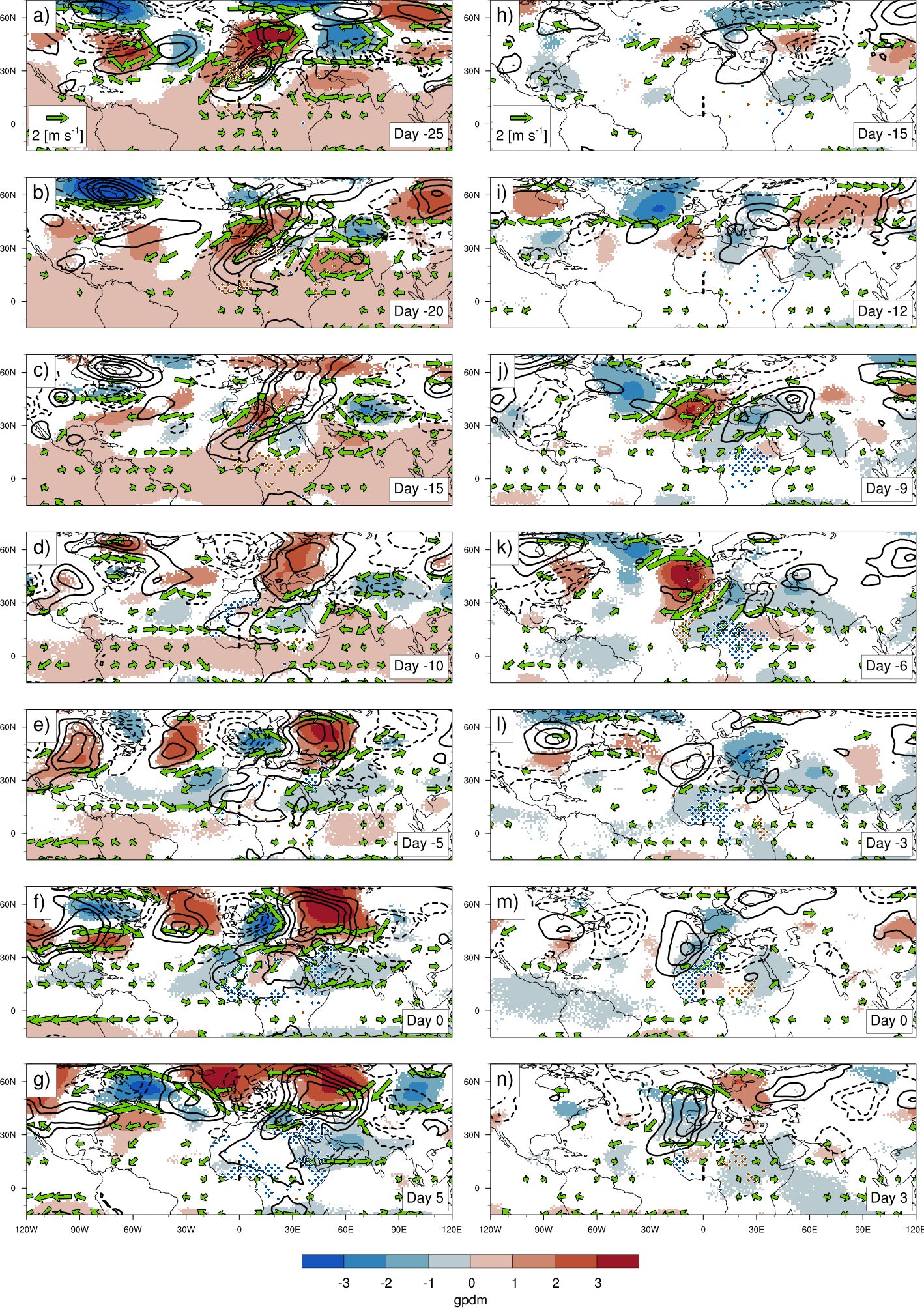}\\
		\caption{Time lag composites of wind (vector) and geopotential (shading) in \SI{300}{\hecto\pascal} and geopotential thickness between 925 and \SI{600}{\hecto\pascal} (contours, from -1 to 1 gpdm in 0.25 steps, negative dashed) for days before the occurrence of a tropical plume-like MJO signal during phase 4 (a-g) and an ER signal during phase 6 (h-n). Significant precipitation anomalies over Africa are shown with blue (positive) and red (negative) dots. The wave signal was filtered for \ang{5}--\ang{15}N, \ang{0}E (dashed line). The analyzed time period spans from 1979 to 2013 for the extended monsoon season (March--October).}
		\label{fig.LagMJO_ER}
	\end{figure*}

	The MJO couples with extra-tropcial Rossby waves as the associated precipitation and circulation patterns suggest, which resemble tropical plumes (Figs.~\ref{fig.mapMJO}d and S1d). During phase 4, precipitation is enhanced south of a trough located over the Atlas mountains. The time lag composites prior to the tropical plume show that positive geopotential height anomalies are evident in the tropics one month prior to rainfall events (Fig.~\ref{fig.LagMJO_ER}a). In the Northern Hemisphere, a quasi-stationary Rossby wave train exists, which moves slowly westward. The composites suggest that the MJO band rather reflects a standing than an eastward propagating mode. A dry anomaly persists over Africa for several weeks and vanishes as soon as the geopotential anomaly vanishes. Related to the Rossby wave train over the Atlantic Ocean, the SHL is strengthened ahead of a rigde over Central Europe (Figs.~\ref{fig.LagMJO_ER}a-c). Following this rigde, a trough amplifies over Central Europe ten days later, which merges with a trough over the Atlas mountains (Figs.~\ref{fig.LagMJO_ER}e-f). The related south-westerlies over the Sahara transport moisture northwards and facilitate the generation of rainfall over the region (Fig.~\ref{fig.mapMJO}d). Additionally, to the geopotential anomalies over the Atlantic Ocean and Europe, a cyclonic and anti-cyclonic vortex pair exist over the Indian subcontinent 25--15 days before the event (Figs.~\ref{fig.LagMJO_ER}a-c). This indicates a potential influence of the Indian monsoon system on the WAM on the intraseasonal timescale.
	
\begin{figure}
		\centering
		\noindent\includegraphics[clip, trim=7cm 11cm 7cm 11cm, width=5cm]{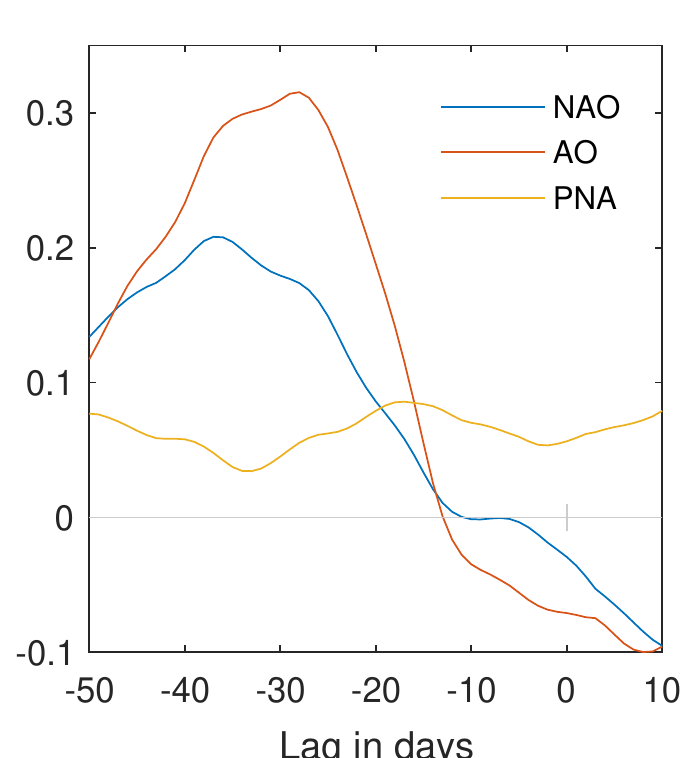}\\
		\caption{Time lag analysis of mean NAO, AO, and PNA indices before significant MJO signals in local phase 4 at \ang{0}E. A 15-day running mean was applied to the timeseries in order to highlight the intraseasonal variability.}
		\label{fig.indices}
	\end{figure}

	The quasi-stationarity of the signal in the extratropics suggests a possible link to other intraseasonal modes. The geopotential patterns at day -25 resemble a positive NAO pattern that reverses until day 0. Although, the AO is weak during boreal summer, a positive AO precedes the event, as negative geopotential anomalies in polar and positive anomalies in subtropical regions suggest. Time lag analysis of the NAO, AO, and PNA indices reveal that positive NAO and AO phases are more likely 40 to 20 days before the plume (Fig.~\ref{fig.indices}). A strong connection to the PNA is not likely, as the index is weakly positive during the time before and shortly after the event. 
	
	ER waves also show precipitation pattern over the Sahara reminiscent of tropical plumes. It seems from Fig.~S2 that the westward propagating ER signal interacts with the eastward movement of extratropical Rossby waves, triggering a wave breaking and leading to a tropical plume in phase 6 (Fig.~\ref{fig.mapER}f). Enhanced precipitation can then be found south and south-east of a trough over the Atlas mountains. Twelve days before the event, weak geopotential anomaly patterns reach from the Atlantic Ocean to the Indian subcontinent, forming a Rossby wave train (Fig.~\ref{fig.LagMJO_ER}i). As it intensifies, a ridge over western Europe triggers cold south-easterlies over northern Africa, which ventilate the northern part of the SHL. Over northern India, significant geopotential and thickness anomalies are evident at day -15 (Fig.~\ref{fig.LagMJO_ER}h). These anomalies travel westwards and intensify over Central and East Africa (Figs.~\ref{fig.LagMJO_ER}j-k). A trough, which stretches from the Atlas mountains to Central Europe, strengthens and triggers north-westerly wind anomalies that subsequently lead to the formation of the tropical plume (Figs.~\ref{fig.LagMJO_ER}l-m). Based on the time-lagged composites we hypothesize that breaking extratropical Rossby wave can project onto the ER band. It should also be noted that the observed geopotential height patterns do not show a pure symmetric ER signal as expected from theory. 
	
	%%%%%%%%%%%%%%%%%%%%%%%%%%%%%%%%%%%%%%%%%%%%%%%%%%%%%%%%%%%%%%%%%%%%%
	% Discussion
	%%%%%%%%%%%%%%%%%%%%%%%%%%%%%%%%%%%%%%%%%%%%%%%%%%%%%%%%%%%%%%%%%%%%%
	
	\section{Discussion}  \label{sec:Discussion}
	This section will relate the results presented in the previous section with former research. The influence of tropical waves on circulation patterns and moisture modulation will be examined. Then, the difference in the vertical structure and implications for rainfall modulation will be discussed. Finally, the relationship of MJO and ER waves with the extratropics will be reviewed.
	
	\subsection{Circulation patterns and moisture modulation}
	As shown in the previous section, specific circulation patterns are associated with each wave type. Inertio-gravity waves are more divergent whereas ER waves are predominantly rotational \citep{Delayen.2016}. Kelvin waves are mostly divergent modes, although at low wavenumbers ($k<5$) they transition to a more rotational behavior. Westward MRG waves behave similar to ER waves at high wavenumbers, but exhibit more divergent characteristics at low wavenumbers. ER (n=1) and Kelvin waves are symmetric about the equator, whereas MRG and EIG (n=0) waves are antisymmetric \citep{Matsuno.1966}. Circulation patterns of these waves over the study area show properties similar to the theoretical solutions. The symmetric or antisymmetric counterparts in the Southern Hemisphere are weaker, though. In the following, results from previous research on single waves will be compared to our results.
	
	For the MJO, \cite{Matthews.2004}, \cite{Janicot.2009}, and \cite{Alaka.2012} show that the effect over Africa is a result of Kelvin and Rossby response to enhanced convection over the warm pool. Thus, the MJO shares several properties of Kelvin and Rossby waves such as the strong modulation of zonal winds and vortices off the equator. \citet{Matthews.2004}, and  \citet{Alaka.2012} also describe increased moisture supply as shown in this study. \citet{Lavender.2009} argue, though, that the increased boundary layer moisture is rather a result of enhanced convection than a cause, which is consistent with our results. The increased cyclonic shear at the southern flank of the AEJ during the moist phase of the MJO is also mentioned by \citet{Matthews.2004}, and \citet{Ventrice.2011}. As a side remark, the increased horizontal shear affects the stability of the AEJ and thus the AEW occurence is modified (see also \citeauthor{Alaka.2012} 2012; 2014).
	
	\citet{Janicot.2010} study in detail ER waves and their impact on the WAM. They find associated circulation fields that are similar with the results of this study. Additionally, they show that the 10--30 days ER mode are likely part of the so called "Sahel-mode" \citep{Sultan.2003b}. The results presented here are also comparable with the composites presented by \citet{Thiawa.2017}. As both slowly propagating modes, the ER waves and the MJO, strongly modulate PW, they are also the only wave types significantly shifting the ITD position by up to 100--\SI{200}{\kilo\meter}. The triggering of tropical-plume like rainfall anomalies over the Sahara and the relationship to the extratropics will be discussed in the following sub-section \ref{sec:dis_rossby}.
	
	MRG waves over the African continent are not well studied. As MRG waves have a strong meridional component at the equator, they strongly affect monsoonal inflow. MRG waves share several similarities with TDs. As their spectral bands partially overlaps, both modes can project onto each other. Over the Pacific, MRG waves can transition to off-equatorial TDs \citet{Takayabu.1993}. A close interaction between both wave modes is also suggested in \citet{Cheng.2018}.
	
	The circulation patterns of Kelvin waves shown in Fig.~\ref{fig.mapKelvin} are consistent with previous studies over Africa \citep{Mounier.2007,Nguyen.2008,Sinclaire.2015,Thiawa.2017}. Kelvin waves are commonly described as modes that only affect zonal winds. Notably, theoretical studies have revealed that Kelvin waves exhibit additional meridional circulation if the forcing lies off the equator \citep{Gill.1980, Dias.2009}. Consistent with observations for equatorial Africa \citep{Mounier.2007,Sinclaire.2015}, this study confirms that Kelvin waves also affect northerly moisture influx during the wet phase of the wave (Fig.~\ref{fig.mapKelvin}).
	
	TDs correspond mainly to AEWs over Africa \citep{Roundy.2004}. Thus, the circulation patterns of TDs presented here match well with the structure of AEWs. In the region of northerlies, precipitation is enhanced \citep{Duvel.1990, Fink.2003,Janiga.2016}. A secondary maximum, which can be found in the regions of southerlies, is not resolved by the filtering method applied here. As previously mentioned, the TD band might be also partially contaminated by MRG or even ER signals, which can influence the TD structure \cite{Yang.2018}.
 
	The role of EIG waves on the WAM has not been studied so far. This study shows that the influence of EIG waves on the dynamics and thermodynamics of the WAM are minor. As mentioned by \cite{Kiladis.2016}, EIG waves are more difficult to capture due to their weak influence on circulation. A drawback of the applied methods is that the filtering, which is based on daily OLR measurements, might miss the fast EIG signal. Additionally, reanalysis might not adequately capture these imbalanced modes, which are to a large extent not assimilated by current NWP models \citep{Zagar.2016}. 
	
	\subsection{Vertical structure and mechanism of rainfall modulation}
	
	All tropical waves, except ER waves, have a self-similar vertical structure \citep{Mapes.2006,Kiladis.2009}. CCEWs are tilted vertically. Ahead of the region of enhanced rainfall, winds converge and moisture is increased at low levels. Mid levels moisten during the progression of the wave until the entire troposphere is moist during the deep convective phase. Furthermore, \citet{Roundy.2012} predict theoretically such a behavior for imbalanced waves using the Boussinesq model. This model is an alternative theory to the shallow-water model, which additionally takes buoyancy into account. In balanced waves, buoyancy does not act as restoring force and thus they do not possess such a vertical tilt. The radiosonde analysis confirms these results. The more imbalanced modes (Kelvin, TD, and MRG) show a tilted vertical structure, whereas the balanced ER wave lacks such a tilt. This agrees with observations by \cite{Yang.2018}. Interestingly, they show that this tilt depends on longitude and is the strongest over West Africa. To test this, was beyond the scope of this study. The vertical structure of the MJO, which was not analyzed in this study, is similar to a Kelvin wave \citep{Mapes.2006,Tian.2006,Roundy.2012b,Jiang.2015}. 
	
	Rainfall anomalies in the more balanced ER wave are mostly related to large-scale moistening and quasi-geostropic lifting. Stratiform rain is thus facilitated. The rainfall anomaly in predominantly imbalanced waves (MJO, Kelvin, and MRG) is of convective nature. Increased CAPE and low-level shear as well as decreased mid-level humidity favor vigorous and organized convection during the wet phase. TDs show the same effects. As MRG has imbalanced characteristics at low wavenumbers \citep{Delayen.2016}, the organization of precipitation is not as strongly modulated, compared with the MJO, Kelvin wave, and TD. Several studies have documented that favorable conditions for organized convection exist over Africa during the passage of AEWs \citep[e.g.][]{Duvel.1990,Fink.2003,Janiga.2016,Maranan.2018} and Kelvin waves \citep[e.g.][]{Mounier.2007,Nguyen.2008,Laing.2011,Sinclaire.2015}. The effect of MRG and Kelvin waves on organization was documented by \citet{Holder.2008}. Although EIG waves are by nature imbalanced modes, this study could only reveal a minor influence on vertical profiles. Yet, it could be speculated that the applied methods could not measure this adequately due to the high speed of EIG waves. 
	
	\subsection{Relationship with the extratropics}
	\label{sec:dis_rossby}
	The MJO over northern tropical Africa can trigger slightly poleward and eastward tilted tropical plumes, which are related to a quasi-stationary signal in the extratropics. The relationship between the tropical and extratropical regime on the intraseasonal timescale is complex. One mechanism of how the MJO influences remote area are through the excitation of Rossby wave trains \citep{Opsteegh.1980,Hoskins.1981,Seo.2017}. This study documents a blocking over Europe 20 days prior to the tropical plume, which is consistent with \citet{Henderson.2016}, who found a significant modulation of blocking frequency over the Atlantic Ocean and Europe. \citet{Alaka.2014} suspected a link of the intraseasonal variability over West Africa with the NAO. This study attests that a positive NAO and AO precede the MJO induced plumes over Africa. A feedback between the MJO and NAO is documented by \citet{Cassou.2008}, and \citet{Lin.2009}. The AO, which is correlated with the NAO, has been also associated with the MJO \citep{Zhou.2005,LHeureux.2008}. A connection with the PNA could not be found in this study, as described for the global MJO by e.g. \citet{Ferranti.1990, Mori.2008, Lukens.2017}, albeit for the boreal winter, whereas this study focuses on the pre-monsoon. It has to be noted also that the observed MJO pattern over Africa has a more standing component (see Fig.~\ref{fig.LagMJO_ER} and \citealt{Pohl.2009}) . Roughly \SI{40}{\percent} of significant local MJO events are missed by the global RMM index (Fig.~\ref{fig.RMM_P4}). Thus, a direct comparison with global MJO studies is difficult. Additional to extratropical influences, the Indian Monsoon system can influence the MJO signal over Africa during the monsoon season \citep{Berhane.2015}. \cite{Janicot.2009} analyzed the intraseasonal variability of West African rainfall and showed that active and break phases of the Indian monsoon also project onto the intraseasonal signal over West Africa 15--20 days later. Figure~\ref{fig.LagMJO_ER}a also suggests such a link to the Indian monsoon. 
	
	ER waves act on a shorter timescale than the MJO. The involved mechanism is likely different. A similar patterns of ER waves showing a trough over the Atlas mountains was documented also by \cite{Janicot.2010} and \citet{Thiawa.2017}. Figure~\ref{fig.LagMJO_ER} suggests that this trough is related to a Rossby wave train over the Atlantic. During the pre-monsoon, when most of the cases happen, the Rossby wave train can penetrate far south, as the STJ is still located at \ang{25}N. The breaking of these Rossby waves leads to a tilted trough over the Atlas mountains and subsequently triggers a tropical plume similarly to the process described by \cite{Knippertz.2003, Knippertz.2005}, and \cite{Frohlich.2013}. From the time-lag analysis, it is not clear, however, to which portion ER waves can be attributed, i.e. to an extratropical wave train or to tropical origin. Further research is needed here. Separating both regimes would help to answer this question. As a final side remark, \citet{Vizy.2009, Vizy.2014, Chauvin.2010} and \citet{Leroux.2011} documented a similar mechanism how a mid-latitude wave train originating over the Atlantic can influence the intraseasonal variability of the WAM. \citet{Chauvin.2010,Leroux.2011} and \cite{Roehrig.2011} point to the role of the SHL, which serves as a link between the extratropics and tropics. Consequently, the strength and location of the SHL affects the activity of AEW and intraseasonal variability of rainfall over northern tropical Africa. These studies and the results presented here emphasize the relevance of extratropical forcing for the intraseasonal rainfall variability in the study region.
	
	%%%%%%%%%%%%%%%%%%%%%%%%%%%%%%%%%%%%%%%%%%%%%%%%%%%%%%%%%%%%%%%%%%%%%
	% Conclusion
	%%%%%%%%%%%%%%%%%%%%%%%%%%%%%%%%%%%%%%%%%%%%%%%%%%%%%%%%%%%%%%%%%%%%%
	
	\section{Conclusion}  \label{sec:Summary}
	Precipitation and the necessary dynamic and thermodynamic conditions are known to vary systematically in space and time in tropical regions like Africa. Tropical waves are the main factor determining the atmospheric environmental conditions that facilitate or suppress precipitation on daily to monthly timescales. So far no systematic comparison has been performed to analyze these waves and their influence on rainfall, dynamics, and thermodynamics over the region of northern tropical Africa. The present study complements SFKV18, who quantified the influence of tropical waves on rainfall over the region. This part investigated the effect on the dynamics and thermodynamics within the WAM. The key results are:
	
	\begin{itemize}	
		\item Tropical waves show specific circulation patterns that are largely consistent with theoretical predictions. The modulation by EIG waves is very weak. The slow modes, MJO and ER waves, have a strong impact on PW, whereas moisture convergence is the dominant factor for rainfall generation in the faster Kelvin, and MRG waves, as well as TDs. Monsoonal inflow is increased during wet phases of the MJO, and ER and MRG waves. Due to the slow speed of the MJO and ER waves and their influence on PW, both waves modulate the zonal position of inter-tropical discontinuity by about 100 to 200 km.
		\item Radiosonde data reveal the vertical tilt of imbalanced wave modes (Kelvin, MRG waves and TDs). The balanced ER waves are not vertically tilted. Organization of deep convection is facilitated during the wet phases of the imbalanced modes, MJO, Kelvin waves and TDs. Slightly favorable conditions occur during the passage of MRG waves. Organization is hampered during the passage of the balanced ER mode. Rainfall triggered by ER waves is more likely generated by large moistening and stratiform lifting.
		\item The MJO and ER waves interact with the extratropics. A time-lag analysis suggests that in both cases an extratropical Rossby wave train causes a trough over the Atlas mountains, which then triggers rainfall over the Sahara. Additional influence stems from intraseasonal variability of the Indian monsoon system. The extratropical Rossby wave signal is likely the result of the MJO, whereas the ER signal is partially triggered by a related extratropical Rossby wave train over the Atlantic and partially has tropical origin. Time-lag analysis indicates that the MJO event is likely preceded by a positive NAO and AO signals.
	\end{itemize}
	
	Several scientific questions remain open: This study highlighted the differences between the effect of imbalanced modes in the WAM region compared to balanced modes. Current global NWP models do not assimilate imbalanced modes. This stresses the need of new schemes such as those proposed by \citeauthor{Zagar.2005} (\citeyear{Zagar.2005,Zagar.2012,Zagar.2016}). This study only analyzed thermodynamic variables and did not consider other diabatic fluxes, such as radiation. More research is needed to investigate whether radiative fluxes play a significant role in the thermodynamics of tropical waves. Finally, the teleconnection patterns for the MJO and ER wave documented here should be analyzed in more detail because of their potential as intraseasonal predictors for Sahelian rainfall during the pre-monsoon. 
	
	Future developments of rainfall prediction models can benefit from the results presented here. This study unveil which dynamical processes need to be modeled realistically to represent the coupling between tropical waves and rainfall. Operational NWP forecasts are improving in the representation of tropical waves \citep{Janiga.2018}. In particular, the large-scale modulation of winds and moisture by the MJO, the quasi-geostrophic equatorial (and related extratropical) Rossby waves are predicted reasonably well on a timescale of two to three weeks \citep{Grazzini.2015,Li.2015,Janiga.2018,Tseng.2018}. However, as numerical models still struggle to represent mesoscale structures of organized convection, the modulating influence of smaller scale imbalanced tropical waves on rainfall might be missed. This gap could be addressed by developing dynamical-statistical models that incorporate real-time filtered wave information from NWP models and observations (\citealt{Wheeler.2001,Wheeler.2004,Roundy.2009,Janiga.2018}) as predictors for rainfall, potentially leading to better rainfall predictions over northern tropical Africa.

	\textbf{Acknowledgment}
	
	\footnotesize
	The research leading to these results has been accomplished within project C2 ``Prediction of wet and dry periods of the West African Monsoon'' of the Transregional Collaborative Research Center SFB / TRR 165 ``Waves to Weather'' funded by the German Science Foundation (DFG).  We thank Tilmann Gneiting and Peter Vogel for discussions and comments on draft versions of the paper. The authors also thank various colleagues and weather services that have over the years contributed to the enrichment of the KASS-D database; special thanks go to Robert Redl and Benedikt Heyl for creating the IGRA2reader package. The authors also thank Roderick van der Linden for providing code that was further developed to filter the waves and create the composite plots. 
	
	\bibliography{references}

\onecolumn

\renewcommand{\thefigure}{S\arabic{figure}}
\setcounter{figure}{0} 

\center \large \textbf{Supplementary Material}
		
\begin{figure}[h!]
	\centering
	\noindent\includegraphics[height = 0.84 \textheight ]{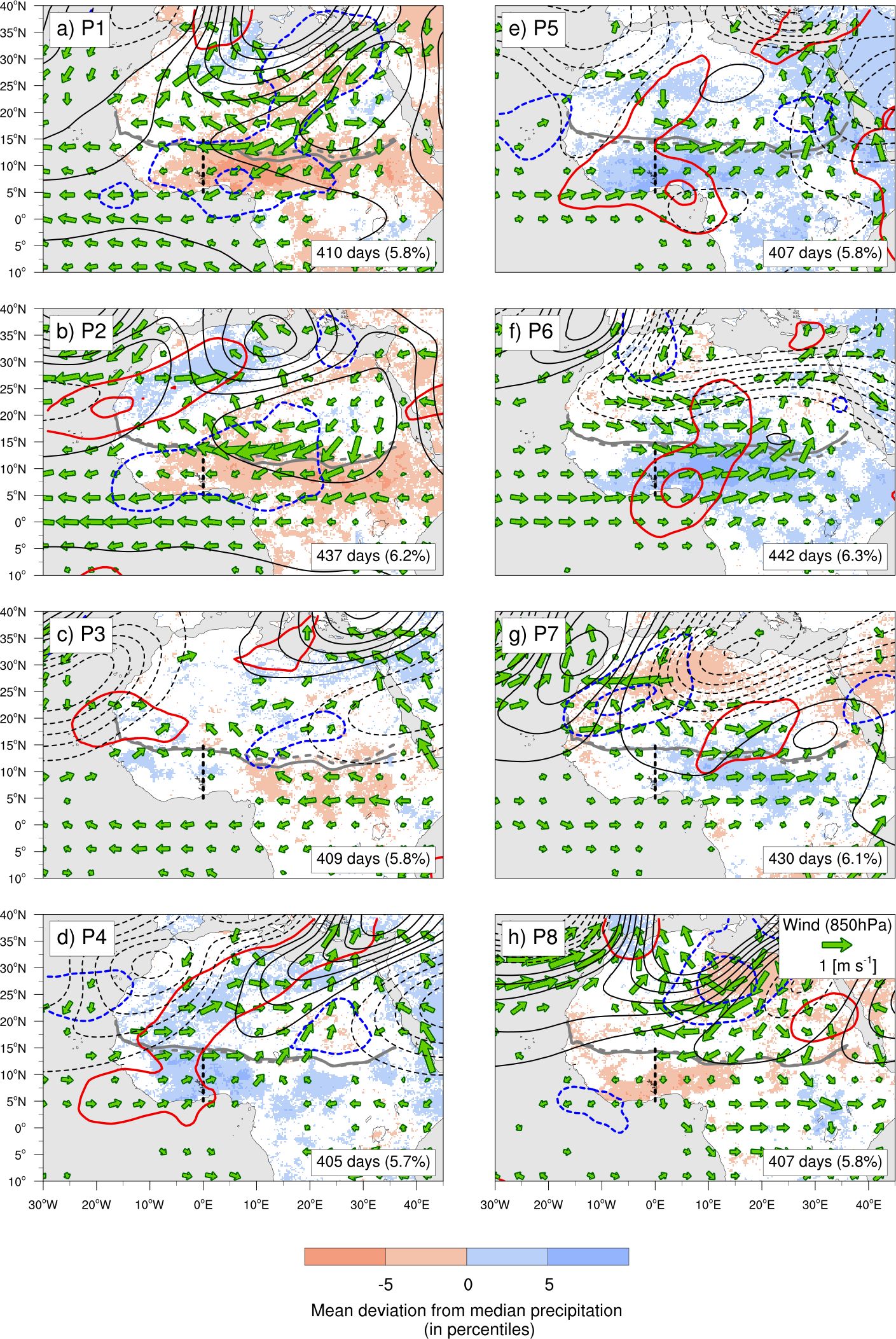}\\
	\caption{Similar composite as Fig.~1 in the main paper, but showing significant \SI{850}{\hecto\pascal} wind anomalies (green arrows), precipitation anomalies (shaded), convergence in \SI{200}{\hecto\pascal} (thick contours, at \SI{-3E-6}{\per\second} to \SI{3E-6}{\per\second} in \SI{0.5E-6}{\per\second} steps, positive solid red lines, negative dashed blue lines), and geopotential height anomalies in \SI{300}{\hecto\pascal} (thin black contours, at -1.4 to 1.4 gpdm in 0.2 gpdm steps, negative dashed) for the Madden-Julian Oscillation. Gray line indicates the ITD position.}\label{fig.S_mapMJO}
\end{figure}

\begin{figure*}[p]
	\centering
	\noindent\includegraphics[height = 0.95 \textheight ]{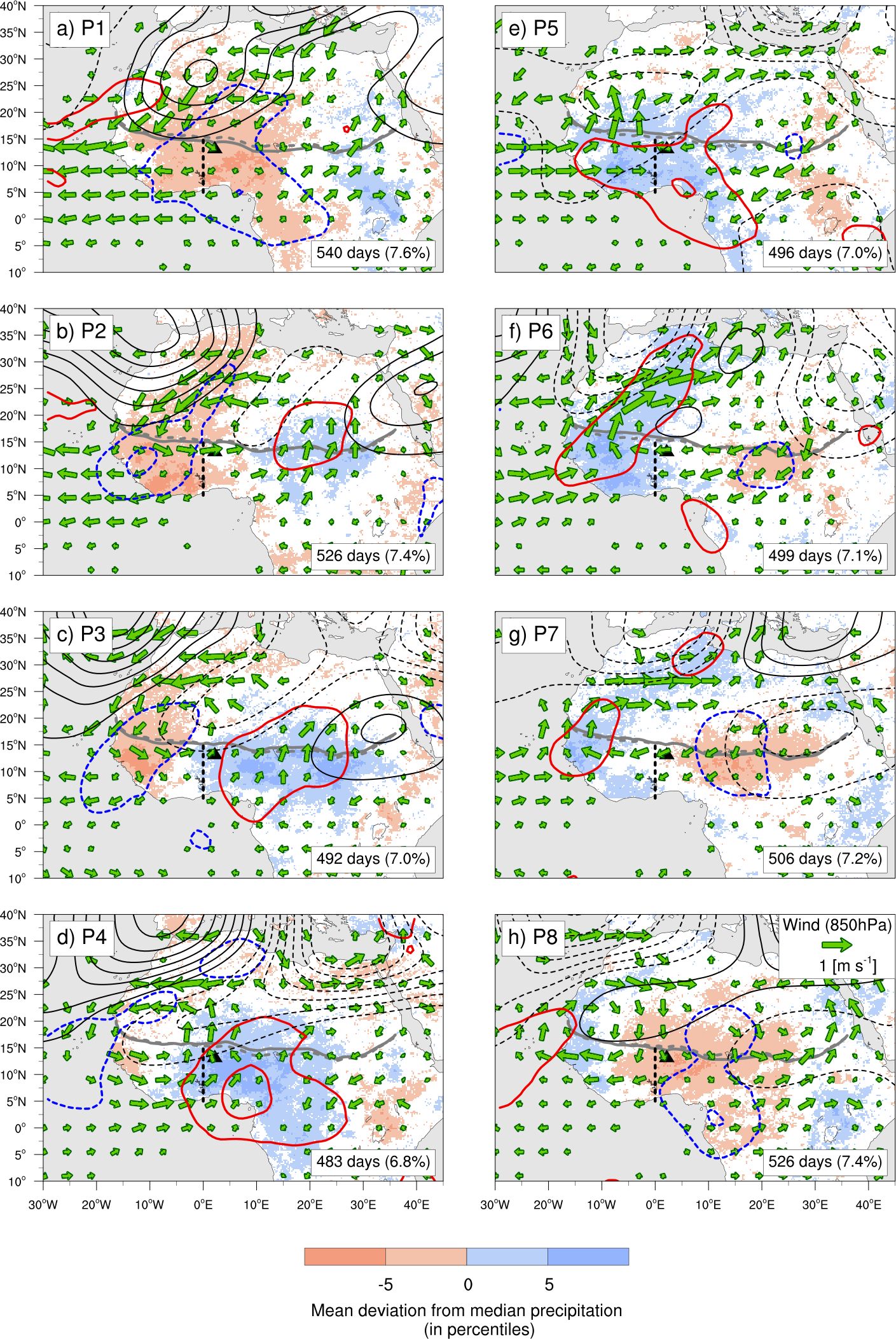}\\
	\caption{Same as Fig.~\ref{fig.S_mapMJO}, but for equatorial Rossby waves.}
\end{figure*}

\begin{figure*}[p]
	\centering
	\noindent\includegraphics[height = 0.95 \textheight ]{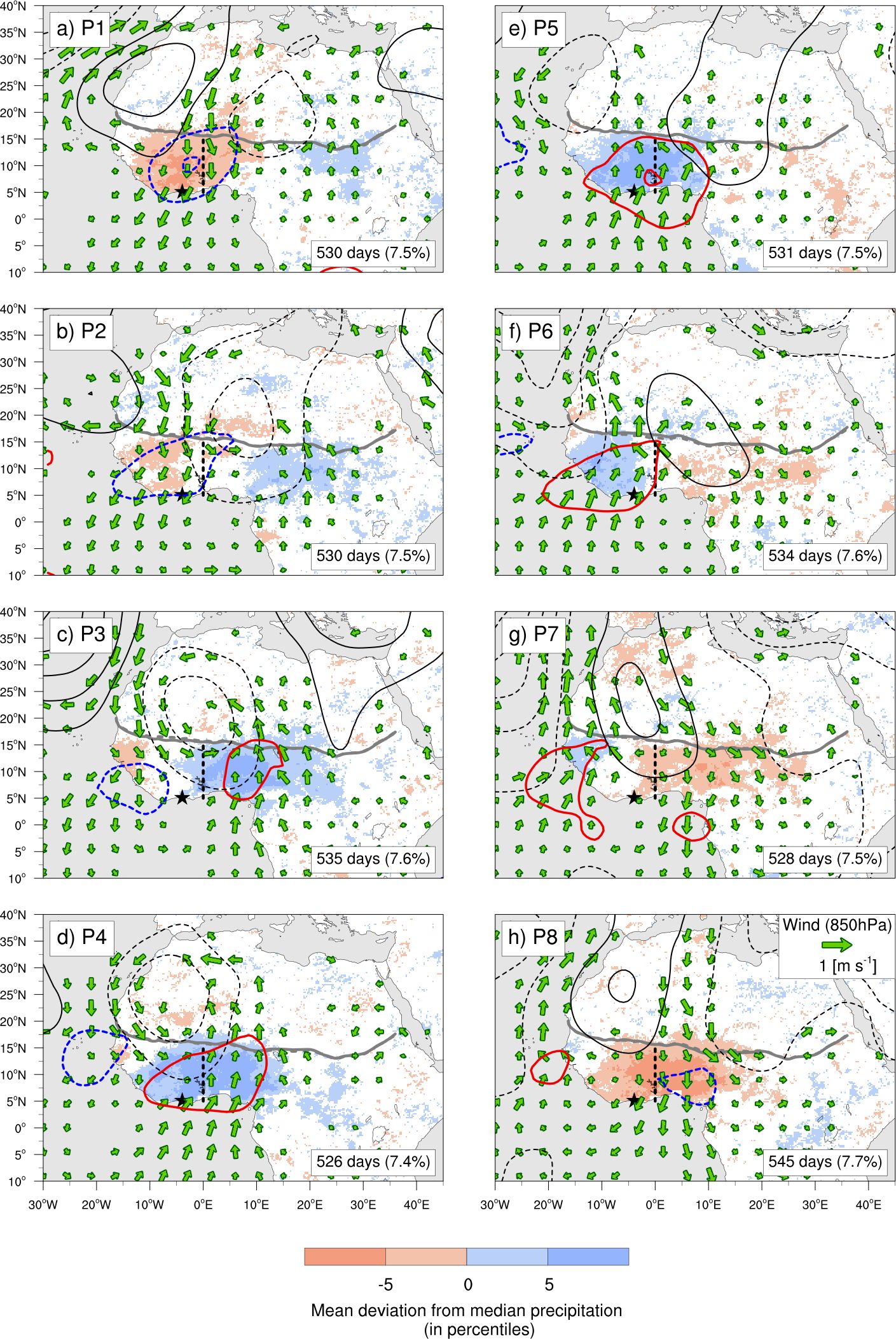}\\
	\caption{Same as Fig.~\ref{fig.S_mapMJO}, but for mixed Rossby gravity waves. Thin black contours show \SI{850}{\hecto\pascal} geopotential height anomalies at -0.5 to 0.5 gpdm in 0.1 gpdm steps (negative dashed).}
	\label{fig.S_mapMRG}
\end{figure*}

\begin{figure*}[p]
	\centering
	\noindent\includegraphics[height = 0.95 \textheight ]{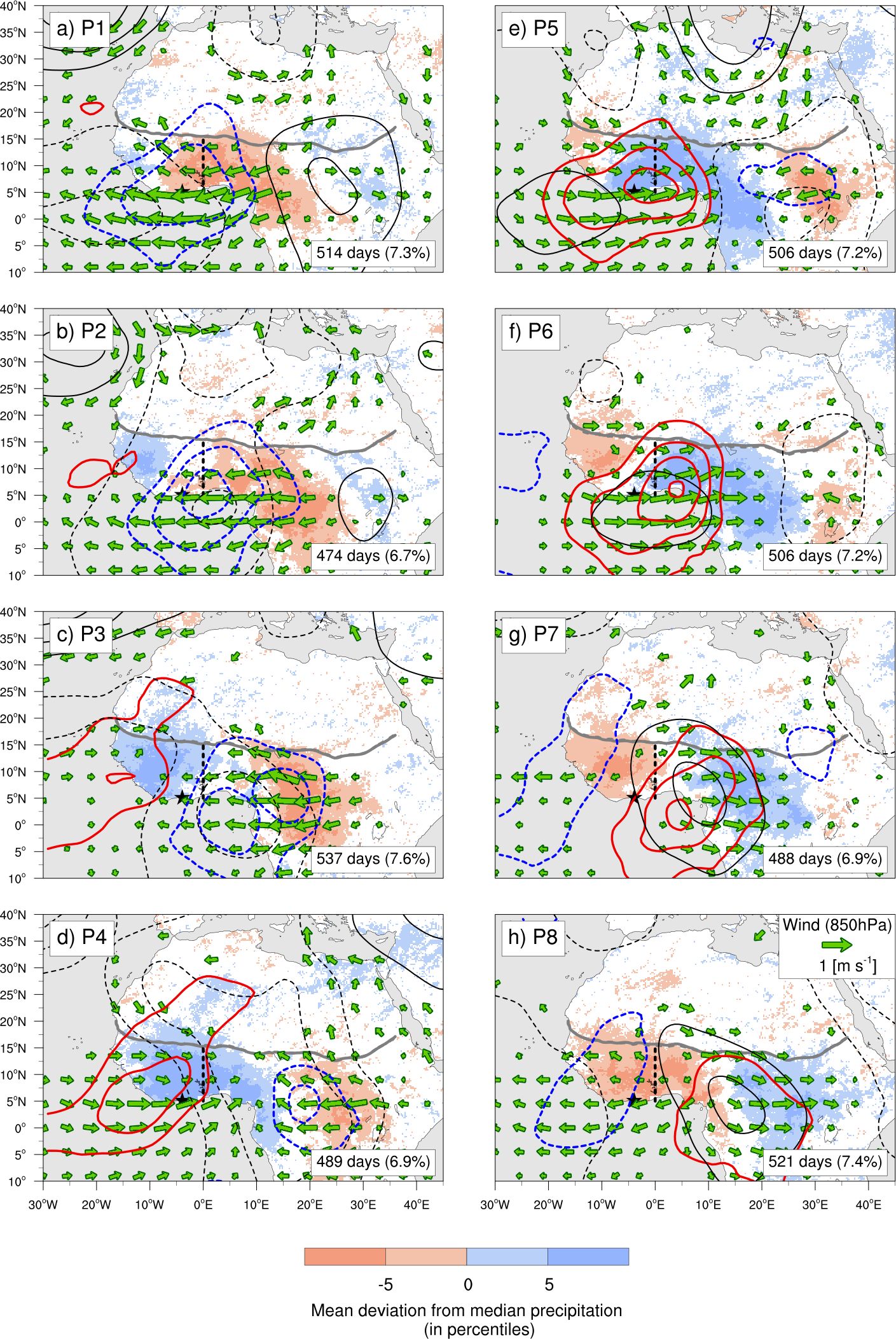}\\
	\caption{Same as Fig.~\ref{fig.S_mapMRG}, but for Kelvin waves.}
\end{figure*}

\begin{figure*}[p]
	\centering
	\noindent\includegraphics[height = 0.95 \textheight ]{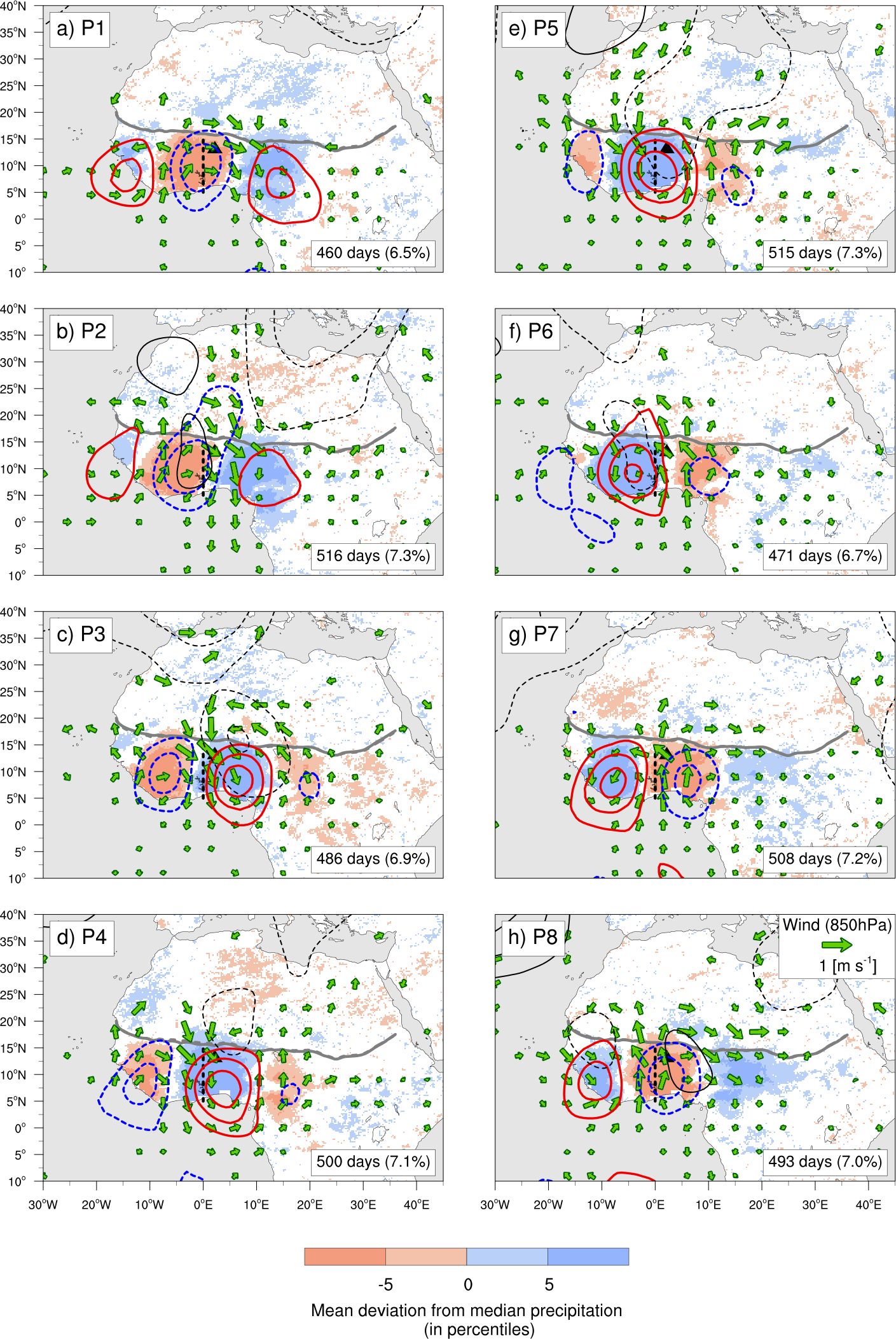}\\
	\caption{Same as Fig.~\ref{fig.S_mapMRG}, but for tropical disturbances (mostly African Easterly Waves).}
\end{figure*}

\begin{figure*}[p]
	\centering
	\noindent\includegraphics[height = 0.95 \textheight ]{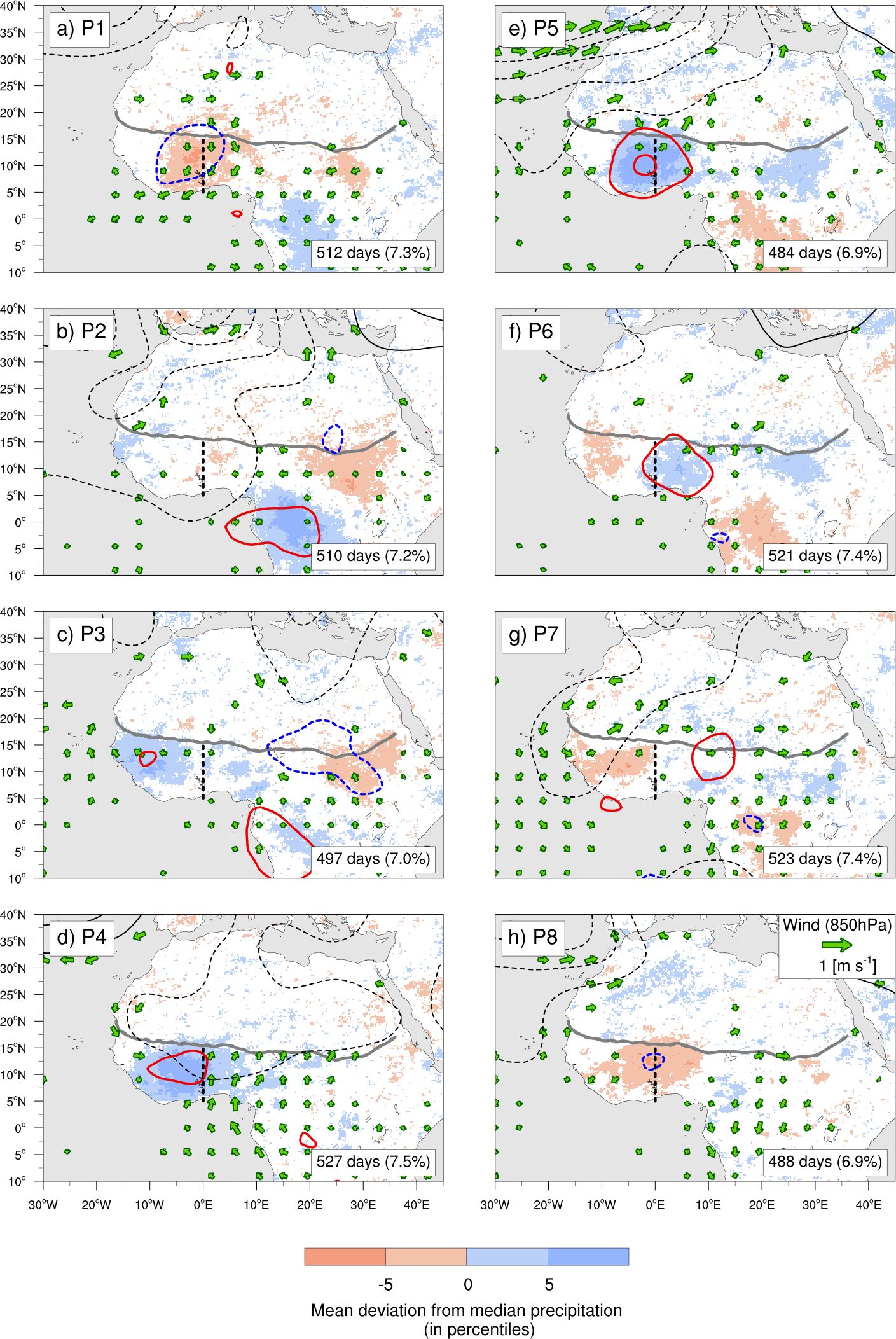}\\
	\caption{Same as Fig.~\ref{fig.S_mapMRG}, but for eastward inertio-gravity waves.}
\end{figure*}

\begin{figure*}[p]
	\centering
	\noindent\includegraphics[height = 0.95 \textheight ]{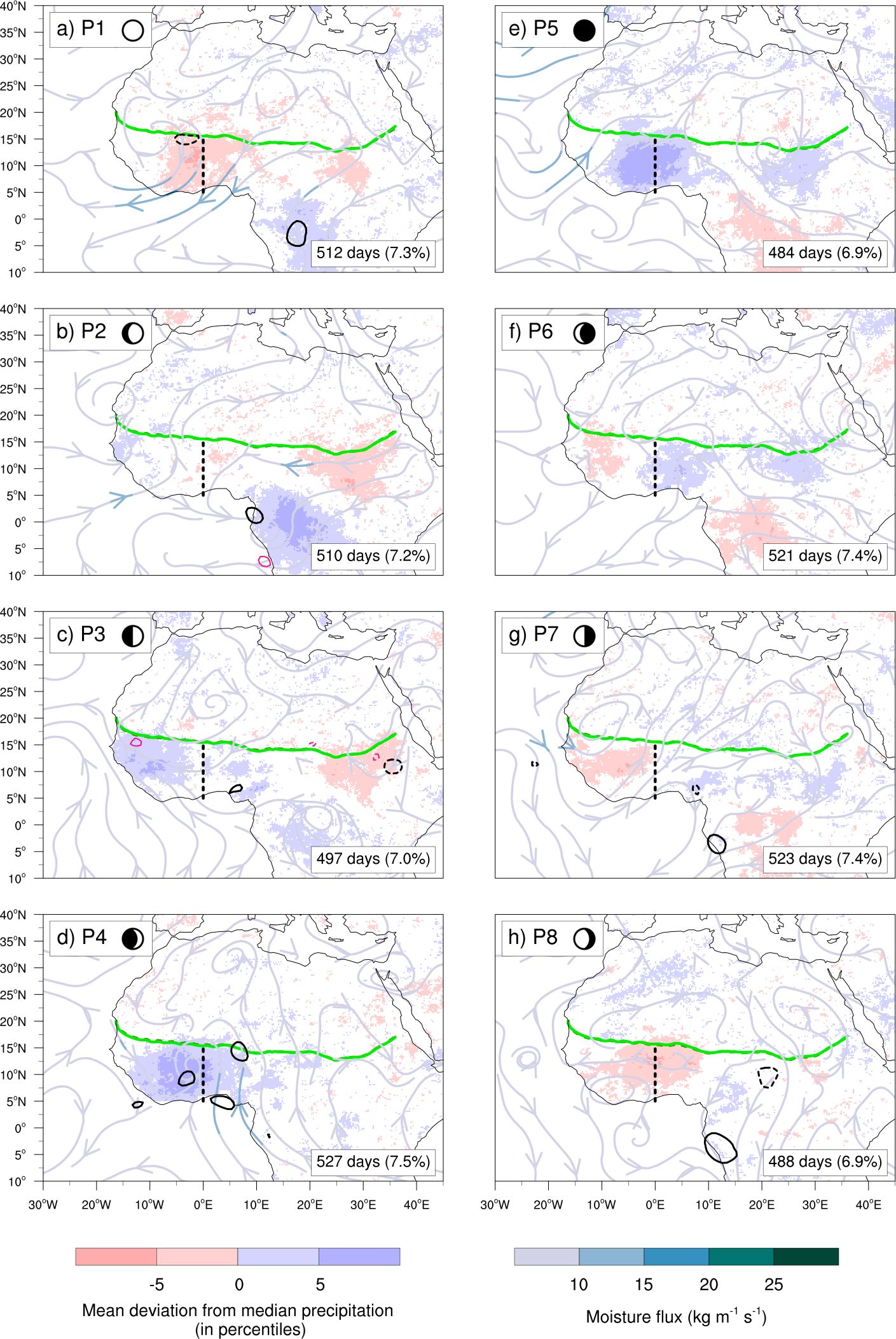}\\
	\caption{Same as Fig.~1 in the main paper, but for eastward inertio-gravity waves.}
\end{figure*}
		
\end{document}